\documentclass[sigconf]{acmart}
\AtBeginDocument{%
  }

\copyrightyear{2026}
\acmYear{2026}
\setcopyright{cc}
\setcctype{by}
\acmConference[CHI '26]{Proceedings of the 2026 CHI Conference on Human Factors in Computing Systems}{April 13--17, 2026}{Barcelona, Spain}
\acmBooktitle{Proceedings of the 2026 CHI Conference on Human Factors in Computing Systems (CHI '26), April 13--17, 2026, Barcelona, Spain}
\acmPrice{}
\acmDOI{10.1145/3772318.3791427}
\acmISBN{979-8-4007-2278-3/2026/04}

\usepackage{framed}
\usepackage{enumitem}
\usepackage{accents}
\usepackage[inkscapelatex=false]{svg}

\newcommand{\risk}{\textbf{\texttt{Risk}}}    
\newcommand{\ambi}{\textbf{\texttt{Ambiguity}}}

\newcommand{\neutral}{\textbf{\texttt{Neutral}}}    
\newcommand{\ppref}{\textbf{\texttt{Preference Data}}}
\newcommand{\psens}{\textbf{\texttt{Sensitive Data}}}

\begin{document}

\title{The Data-Dollars Tradeoff: Privacy Harms vs. Economic Risk in Personalized AI Adoption}

\author{Alexander Erlei}
\email{alexander.erlei@wiwi.uni-goettingen.de}
\orcid{0000-0001-7322-2761}
\affiliation{%
  \institution{University of Göttingen}
  \city{Göttingen}
  \country{Germany}
}

\author{Tahir Abbas}
\email{tahir.abbas@wur.nl}
\orcid{0000-0002-0558-6106}
\affiliation{%
  \institution{Wageningen University and Research}
  \city{Wageningen}
  \country{The Netherlands}
}

\author{Kilian Bizer}
\email{bizer@wiwi.uni-goettingen.de}
\orcid{0000-0002-2258-7725}
\affiliation{%
  \institution{University of Göttingen}
  \city{Göttingen}
  \country{Germany}
}

\author{Ujwal Gadiraju}
\email{u.k.gadiraju@tudelft.nl}
\orcid{0000-0002-6189-6539}
\affiliation{%
  \institution{Delft University of Technology}
  \city{Delft}
  \country{The Netherlands}
}


\begin{abstract}
Privacy concerns significantly impact AI adoption, yet little is known about how information environments shape user responses to data leak threats. We conducted a 2 $\times$ 3 between-subjects experiment ($N=610$) examining how risk versus ambiguity about privacy leaks affects the adoption of AI personalization. Participants chose between standard and AI-personalized product baskets, with personalization requiring data sharing that could leak to pricing algorithms. Under risk (30\% leak probability), we found no difference in AI adoption between privacy-threatening and neutral conditions (ca. 50\% adoption). Under ambiguity (10-50\% range), privacy threats significantly reduced adoption compared to neutral conditions. This effect holds for sensitive demographic data as well as anonymized preference data. Users systematically over-bid for privacy disclosure labels, suggesting strong demand for transparency institutions. Notably, privacy leak threats did not affect subsequent bargaining behavior with algorithms. Our findings indicate that ambiguity over data leaks, rather than only privacy preferences per se, drives avoidance behavior among users towards personalized AI. 
\end{abstract}




\maketitle

\section{Introduction}
Safety concerns with respect to personal data have long been central for many AI-driven products \citep{windl2022automating,manikonda2018s,menard2025artificial,kokolakis2017privacy}. The current rise of large language models (LLMs) has only amplified these concerns \citep{li2024human}, as people are increasingly motivated to share personal information through conversations \citep{zhang2024s}. Often, people face a trade-off between better services and experiences on the one side, and privacy risks on the other side. Personalization, for example, can improve user experience by providing tailored interactions, while recommendation systems rely on private signals to make calibrated predictions \citep{im2021yes,kocaballi2019personalization,liu2017data,kostric2025should}. Yet, data leaks occur. According to the UK's Cyber Security Breaches Survey, 43\% of all businesses suffered a breach or attack in 2024. The 2025 Thales Data Threat Report states that 45\% of surveyed businesses had experienced a breach at some point. A joint IBM and AWS study asserts that around 76\% of new generative AI products are exposed to privacy and data risks.\footnote{Cyber Security Breaches Survey: \url{https://www.gov.uk/government/statistics/cyber-security-breaches-survey-2025/cyber-security-breaches-survey-2025}; IBM report: \url{https://www.ibm.com/think/insights/generative-ai-security-recommendations}} Notably, data leaks can entail negative economic consequences for users, as other actors may strategically use their data to, e.g., increase prices or target advertisements \citep{aridor2025evaluating,goldfarb2011privacy,braulin2023effects,hidir2021privacy,ali2020voluntary,baik2023price}. For users, these risks are usually opaque, which may deter AI use. As a response, some platforms have started to implement verification systems that improve the information position of consumers. One example are Apple's privacy ``nutrition'' labels \citep{scoccia2022empirical}. However, these labels still suffer from violations and discrepancies between consumer-facing labels and actual policies \citep{ali2023honesty,koch2022keeping}. Consequently, consumers regularly make choices under incomplete information, which inhibits their ability to satisfy preferences and ``reward'' providers that match them \citep{dulleck2006doctors}. In other words, due to the information asymmetry, consumers cannot identify suppliers who fulfill their privacy preferences. Hence, preferences are not communicated via prices, preventing the market from incentivizing privacy safety. This has also been suggested to (partially) explain the gap between peoples' stated privacy preferences and revealed data sharing behavior, i.e. the \textit{privacy paradox} \citep{acquisti2005privacy,buck2014unconscious,baek2014solving}. 

The information environment, meaning how precisely systems communicate privacy information like risk, may thus be a critical determinant in explaining consumer privacy behavior. For most deployed AI systems, users rarely observe the actual probability that their data will be leaked and subsequently misused. Instead, they encounter incomplete provider-information, potentially conflicting signals through, e.g., external media reports, or vague warnings. Furthermore, users will differ substantially in their ability to predict the probability of a data leakage. From a theoretical standpoint, this means that many people often make decisions under \textit{ambiguity}, rather than \textit{risk}. That is, they know that bad outcomes are possible, but cannot attach precise probabilities to them (ambiguity). This distinction matters, because a large body of work in economics and psychology shows that behavior can differ quite substantially between these two information environments. For example, people often react more strongly to ambiguous harms than to risky but well-quantified harms \citep{hogarth1989risk,cohen1987experimental,barham2014roles,dimmock2016ambiguity,anantanasuwong2024ambiguity}. Yet, little is known about how this difference shapes privacy decisions within human-AI interactions, or how interfaces should communicate the corresponding uncertainty to support the informed adoption of AI systems.

Addressing this important gap in privacy research, our work aims to investigate how users respond when confronted with the possibility of a personal data leak in different consumer information environments. Concretely, we (1) examine the behavioral consequences of facing a potential data leakage under different consumer information environments, and (2) empirically disentangle two distinct sources of harm: strategic economic concerns versus non-monetary dis-utility caused by privacy violations. We focus on a 
particularly consequential and underexplored class of privacy harms: personal data leaks that enable third parties to engage in algorithmic price discrimination. This is a particularly intriguing problem, as users face two potential sources of dis-utility following a data leak. One, they may suffer economic costs as sellers are able to secure higher profits through personalized pricing. Two, experiencing a data leak can cause non-monetary harms such as emotional, reputational, or social damage. Despite the significance of these dual effects, which are all the more important given the proliferation of AI systems, prior work has not empirically separated them in the context of privacy-related decision-making.
\footnote{Note that we do not experimentally vary, manipulate, or isolate any specific non-monetary damage from the threat of a privacy leak. Instead, we isolate the economic concerns, and interpret the residual effect as the non-monetary dis-utility caused by facing a potential data leak.} 

We conduct a pre-registered online experiment ($N=610$) using a 2 $\times$ 3 between-subjects design that simulates an e-commerce platform offering an AI personalization option. The AI is economically beneficial, but requires participants to share their personal data, which may later leak to a third party pricing algorithm. We manipulate two factors: the information environment and the mechanism by which future economic costs occur. First, we manipulate whether the probability of the leak is precisely quantifiable (\risk{}) or publicly known to lie within a pre-specified range (\ambi{}). Second, we introduce three data--cost combinations. In \ppref{}, participants share preference data (time preferences, risk preferences, social preferences), and future economic costs from AI personalization are caused by a data leak. In \psens{}, participants share demographic data (gender, age, income range), and future economic costs again arise from a data leak. In \neutral{}, participants share the same demographic data, but future economic costs are caused by a random surcharge lottery that is monetarily equivalent to the data leak. Thus, by holding the expected monetary payoff of using the AI constant between conditions, we isolate privacy-related dis-utility from pure risk and ambiguity aversion.\footnote{Note that we are interested in how the \textit{threat of the leak} itself, as opposed to the potential dis-utility from \textit{sharing} personal data with an AI system, affects AI adoption. Therefore, across all conditions, all subjects must always share their data.}


After the AI personalization decision, participants bargain with an algorithmic seller and state their incentivized willingness to pay for a ``privacy label'' that perfectly reveals data security issues in future AI systems. During the bargaining stage, the algorithm conditions its price-setting on personal information in case of a leak, or charges a one-time surcharge in case of the neutral lottery draw. We address three main research questions:

\begin{framed}
\begin{itemize}[leftmargin=*]
    \item  \textbf{RQ1:} How does the threat of a personal data leak affect AI adoption under known versus ambiguous leak probabilities?
    \item  \textbf{RQ2:} How does the threat of a personal data leak with subsequent economic consequences affect Human-Algorithm bargaining?
    \item  \textbf{RQ3:} To what extent are users and non-users willing to pay for a privacy disclosure label that reveals whether an AI system handles personal data safely?
\end{itemize}    
\end{framed}

Our findings reveal nuanced behavioral responses to privacy risks in AI adoption. When privacy leaks are framed as risky but well-defined, participants exhibit no measurable non-monetary disutility and AI adoption remains stable across conditions at roughly 50\%. 
Under ambiguity, however, people are substantially less likely to use AI personalization in the privacy conditions, and the effect is directionally stronger for \psens{}. Hence, the behavioral consequences of privacy leakage risks are contingent on the informational environment. 
Uncertainty about the likelihood of harm leads users to avoid AI systems that rely on personal data, underscoring the critical role of transparency mechanisms and interface-level communication in supporting informed decision-making.
Incomplete information not only prevents consumers from adopting potentially beneficial technologies, but also 
reduces demand for providers. One promising intervention is the use of third-party verified privacy badges. While costly for providers, our study shows that participants' willingness to pay for such verification closely aligns with theoretical optima, and on average, even exceeds it. 
This contrasts with prior work suggesting low willingness to pay for privacy protections, indicating that verified disclosures may not necessarily impose a competitive disadvantage. 
Finally, we find no compelling evidence that the threat of privacy leaks directly affects consumers' bargaining behavior. Instead, ambiguity about potential price discrimination -- induced by privacy leaks -- makes consumers bargain less aggressively, hurting their economic position. Our findings highlight the importance of how interfaces communicate uncertainty about privacy risks, and the potential of verification mechanisms such as labels or badges to support informed AI adoption without undermining market viability. Our work experimentally disentangles economic and non-monetary harms of privacy leaks in algorithmic price discrimination contexts, reveals how informational risk and ambiguity shapes AI adoption, and offers actionable design and policy implications for transparency and trust in user interactions with AI systems.

\section{Related Work}
A large empirical literature documents that individuals’ stated privacy attitudes often diverge from their actual disclosure and technology-adoption choices, a pattern commonly referred to as the \emph{privacy paradox} \citep{acquisti2015privacy}. Field and lab evidence shows that small frictions, convenience, and contextual cues can induce substantial disclosure even among privacy-concerned users \citep{john2011strangers,beresford2012unwillingness}. More broadly, people weigh perceived benefits against (often opaque or uncertain) privacy costs in a malleable “privacy calculus,” with beliefs, framing, and institutional assurances shaping the trade-off \citep{acquisti2015privacy}. These insights align with classic work on markets with information asymmetries in which consumers struggle to reward high-quality providers when relevant quality dimensions (e.g., privacy, data safety) are unobservable \citep{dulleck2006doctors}.

A first string of empirical papers quantifies the economic side of privacy choices. In a natural experiment on regulatory stringency, \citet{goldfarb2012privacy} show that stronger privacy protection reduced the effectiveness of online advertising, illustrating how limits on personal-data use shift surplus and market outcomes. In purchase settings, randomized field experiments reveal that many consumers are unwilling to pay modest premiums for more privacy-protective options \citep{beresford2012unwillingness}. Furthermore, prior research suggests that participants often demand sizable compensation to reveal sensitive facts while being unwilling to pay comparatively small amounts to protect the same data, consistent with endowment and framing effects in privacy preferences \citep{huberman2005valuating,acquisti2015privacy}. Note that our setting fixes the disclosure of private information between treatments and focuses on (a) the potentiality of a data leak and (b) people's willingness to protect their data in the future. 

A second line of research studies whether transparency and third-party assurances can correct information problems. Providing salient, standardized privacy information (“nutrition labels”) or trust seals can shift behavior toward more privacy-protective choices and higher willingness to transact \citep{tsai2011effect,hui2007value}. At the same time, real-world label ecosystems face credibility and compliance challenges, with documented gaps between disclosures and practices \citep{scoccia2022empirical,ali2023honesty,koch2022keeping}. Our design leverages this tension by offering a perfectly informative label and measuring whether users internalize expected leak costs through their willingness to pay. A related growing experimental literature shows that choice architecture around defaults, recommendations, and interface ordering can generate large differences in disclosure under unchanged objective risk \citep{adjerid2019choice}. On Facebook, simple interface nudges have been shown to measurably alter personal data sharing \citep{wang2014field}. These behavioral patterns imply that observed privacy choices do not necessarily reflect stable preferences or accurate beliefs, pointing towards the information environment (risk. vs. ambiguity) as a potentially influential environmental variable.

The distinction between risky (known probabilities) and ambiguous (incomplete information about the probabilities) information environments has a long tradition in economics and psychology. Research shows that many people exhibit ambiguity aversion, a tendency whereby known-risk choices are preferred over ambiguous ones \citep{fox1995ambiguity,ellsberg1961risk,machina2014ambiguity}. That is, all things equal, people often discount ambiguous options more strongly than risky options. Because uncertainty about the precise likelihood of certain outcomes allows individuals to create subjective probabilities, observed behavior can often be affected by idiosyncratic interpretations which may be difficult to measure, such as optimism, pessimism, or the emotional state \citep{pulford2009short,eichberger2014optimism,baillon2016sadder}. Furthermore, there is ample evidence that individuals try to avoid ambiguity and, therefore, must be paid an ambiguity premium to participate in ambiguous certainty-equivalent lotteries \citep{camerer1992recent}. As AI personalization essentially involves an ambiguous lottery that trades economic gains for a potential data leak, these results suggest a potentially negative effect of incomplete information over the likelihood of a data leak on AI adoption. In the context of privacy, \citet{cao2025privacy} suggest that privacy breaches can drive ambiguity-averse behavior in subsequent decision situations. While our article considers the reverse choice logic where ambiguity over the likelihood of data leakages affects AI adoption, their results indicate a link between privacy concerns, privacy violations, information, and behavior. In contrast to much of the literature on ambiguity aversion, \citet{acquisti2005privacy} document a drop in individuals' revealed valuation of their personal information when they are uncertain about the recipient of their data and the potential use-cases for their data. Here, ambiguity reduces the anticipated costs of a data leak, making AI personalization more attractive. Similarly, \citet{sachs2022privacy} find in an online experiment that subjects prefer to avoid transparency over the likelihood of privacy violations (i.e. risk) and instead seek out ambiguity when privacy violations are related to losses rather than gains. Hence, while the general wisdom over ambiguity preferences should predict stronger aversions towards AI personalization in ambiguous information environments, the empirical evidence in privacy situations, while scarce, points in the opposite direction. Recent work in human-AI collaboration has explored the role of information asymmetry and task uncertainty on AI adoption and human delegation behavior in a variety of contexts~\cite{market4lemons2026,biswas2026belief,salimzadeh2024dealing,he2024err,biswas2025mind}. Interaction with AI and resulting outcomes have been shown to be affected by different human, AI system and task-related factors~\cite{erlei2024understanding,he2023knowing,gadiraju2025enterprising}.  

Finally, work on the personalization–privacy trade-off examines when users accept data sharing in return for better services \citep{asthana2024know}. Early evidence shows that users who demand transparency are often least willing to be profiled for personalization \citep{awad2006personalization}, while perceived high service quality can increase the willingness to share data or pay for personalization \citep{li2012willing}. Prior privacy erosion and perceived sensitivity of the shared attributes can, however, decrease willingness to disclose \citep{li2012willing}. Our experiment mirrors these tensions by contrasting preference data with sensitive demographic data, while holding monetary consequences constant. 

Overall, our study builds on three core aspects of prior literature: (i) privacy behavior is context-dependent and economically consequential \citep{acquisti2015privacy,goldfarb2011privacy,beresford2012unwillingness,huberman2005valuating}; (ii) transparency and credible verification can realign choices but face compliance and design frictions \citep{tsai2011effect,hui2007value,scoccia2022empirical,ali2023honesty,koch2022keeping}; and (iii) the personalization–privacy trade-off is sensitive to uncertainty, ambiguity, data type, and interface framing \citep{awad2006personalization,li2012willing,adjerid2019choice,john2011strangers}. We contribute by causally separating \risk{} from \ambi{} in privacy-leak environments while holding expected monetary outcomes fixed, measuring downstream bargaining behavior with algorithmic sellers, and eliciting incentive-compatible WTP for a perfectly informative disclosure institution that resolves the privacy information problem.

\section{Experimental Design}
\begin{figure}[t]
    \centering
    \includegraphics[width=1\linewidth]{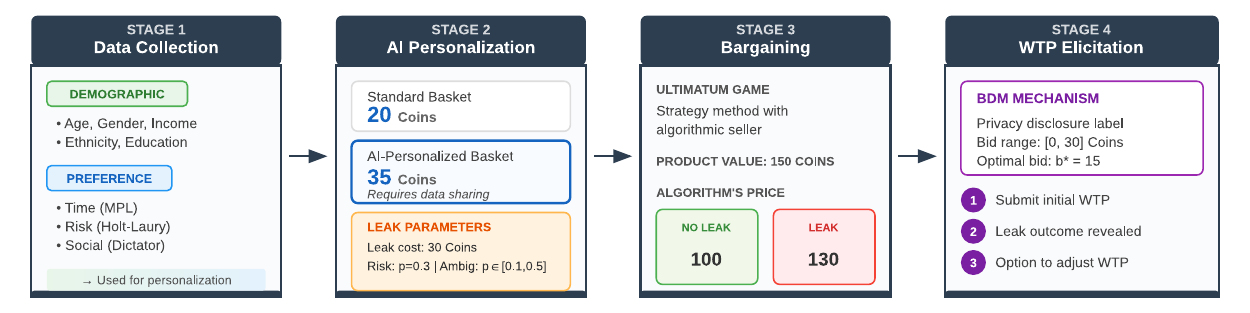}
    \caption{Illustration of the experimental flow, and the four stages that participants completed in our study.}
    \label{exp_flow}
\end{figure}

We deploy a 2 (\risk{} vs. \ambi{}) x 3 (\neutral{} vs. \ppref{} vs. \psens{}) pre-registered between-subjects design.\footnote{Pre-registration: \url{https://osf.io/zwq6d/overview?view_only=619f56a3c2784079a8346499b3617a2d}} The experiment follows a four stage approach as illustrated in Figure~\ref{exp_flow}, which includes (1) personal data elicitation, (2) AI-personalization, (3) bargaining, and (4) willingness-to-pay elicitation. The general procedure is as follows. First, we elicit participants' demographic and preference data. Second, all subjects proceed to an AI-Recommendation stage in which they can choose to utilize a personalized AI assistant for a simple choice task, framed as a product shopping task. Personalization increases the immediate payoff, but may lead to additional costs downstream. In the \textbf{Privacy\_} treatments, using the personalized assistant has a probability $p$ to lead to a subsequent privacy leak, which reduces expected income in the second stage by $z$. In \neutral{}, using the personalized recommendation enters the participant in a lottery that determines with probability $p$ whether any subject must pay $z$ in the second stage for the AI service. In \risk{}, the probability $p = 0.3$ is publicly known, in \ambi{}, subjects know only that $p \in [0.1, 0.5]$.\footnote{Note that this operationalization represents actual market choices when users or consumers face information asymmetries and cannot ex ante observe whether a product stores their private data safely. Most consumers will attach some partially informed prior to that scenario, without having precise knowledge about the overall market-wide risk. Hence, they make choices under ambiguity. More informed individuals, however, may be able to use private expert information to generate more precise priors, and hence, make risky choices.} Finally, in \ppref{}, participants use their preference data (risk preferences, time preferences, social preferences) to personalize the AI, whereas in \psens{}, participants use sensitive demographic data (gender, age, income range), which are protected attributes for non-discrimination according to the GDPR.\footnote{\url{https://gdpr-info.eu/art-9-gdpr/}} Thus, the privacy leak targets either preferences or sensitive personal data, depending on the experimental condition. 

After the AI Recommendation stage, all participants continue to the Bargaining stage. Here, they enter an ultimatum game with an algorithmic price-setting algorithm \citep{guth1982experimental}. The algorithm makes a take-or-leave-it offer on how to split 100 Coins. If the participant rejects, both earn 0, if the participant accepts, they earn the proposed amount. We employ the strategy method. Without a data leak, the seller always proposes a ``fair'' profit split of 50\%, which has been extensively shown to be the welfare-maximizing offer in the absence of any personal information about an individual's preferences \citep{erlei2020impact,erlei2022s,guth2014more}. In case of a data leak, the seller makes a more aggressive offer $50 - z$ Coins, because they can observe the participant's type. This is the economic cost of a personal data leak, reflecting, for example, data-driven personalized price discrimination \citep{ali2020voluntary,dube2023personalized,seele2021mapping}. In the \neutral{} condition, the price increases by $z$ due to the random lottery draw.

Finally, in the fourth stage, subjects state their willingness to pay (WTP) for a privacy label that ensures that they do not experience a data leak in an equivalent second round with a ``new'' platform, a ``new'' AI system and ``new'' sellers. We utilize the BDM elicitation method \citep{becker1964measuring}. For \neutral{}, this is framed as pre-purchasing the personalized AI recommendation service, which, if successful, avoids the lottery. Finally, participants learn whether their data was leaked during the experiment, and then have the option to adjust their prior willingness to pay. Afterwards, the second round starts. The second round shares the same procedure but ends after the bargaining stage, and mainly functions as a real consequence for the WTP stage. 

We pre-registered to collect data until 100 independent observations were gathered per treatment from the crowdsourcing platform CloudResearch Connect \citep{hartman2023introducing}, resulting in a final sample of 610 (42\% female). Participants are English speakers with at least 100 prior tasks on Connect and a minimal approval rating of 90\%. The base payment was \$3.25. Coins earned in the experiment were later converted into Dollars, where 50 Coins = \$1.



\subsection{Procedure: A Detailed Overview}

Here, we provide a more detailed account of the experimental procedure. First, all participants give their consent and proceed to a pre-experimental personal data elicitation stage. They state their age, gender, income range, ethnicity and highest level of education. Then, we elicit time preferences using four multiple price lists (MPL) \citep{coller1999eliciting,andreoni2015measuring} and risk preferences following canonical lottery choices from \citet{holt2002risk}. Subjects complete two lists with 20 choices. The specific parameters for the time and risk preference elicitation tasks are documented in the Appendix. Finally, we capture social preferences using a hypothetical dictator game scenario in which participants allocate \$ 10 between themselves and another anonymous participant. Throughout, we use 2 attention checks and block participants who fail either one from participating.

Afterwards, subjects proceed to the experimental instructions. Here, they learn about the next three stages, the rules of the game, and the payoff scheme. They learn that they will choose one of two product baskets on an e-Commerce platform. To receive the more valuable personalized basket, they must endow the AI system with three private characteristics. In \neutral{}, they provide the demographic data (\psens{}). If they are not willing, they must pick the standard basket. In privacy treatments, we inform participants about the possibility of data leakage and the associated average expected price increase of 30 Coins in Stage 2. They learn that in the event of a data leak, a third party seller's bargaining algorithm will gain access to their data, and use that data when setting prices in Stage 2, leading to the price increase. In \neutral{}, there is no data leak. Instead, subjects learn that using the AI leads to a price increase of 30 Coins with probability $p$ because the bargaining algorithm charges a one-time surcharge if the participant was randomly drawn (with probability $p$) in a lottery that determines who has to pay for the AI's service during the Recommendation Stage. This is analogous to a promotion period in which a supplier wants to attract consumers by randomly selecting some to receive a free product.

Participants must answer four comprehension questions correctly to participate in the experiment. After the instructions, participants complete the AI-Recommendation stage. They can choose between a standard baseline product basket which earns them 20 Coins, or a product basket that has been personalized by an AI system, which earns them 35 Coins. To use the recommender, they must confirm with a mouse-click that they want to submit their data to the AI. They observe exactly what data they submit above the submission button. All subjects are fully informed about the payoffs, price increases, and risky or ambiguous probabilities.

Then, they proceed to the Bargaining stage. We utilize a standard strategy-method ultimatum game setup in which subjects state the minimum offer they are still willing to accept, i.e. in our case, the maximum price they are still willing to pay. This avoids any belief updating regarding the occurrence of a data leak. The game is embedded into the e-Commerce frame as follows. Subjects learn that they want to purchase a product with a value of $x = 150$ Coins. This product is supplied by a seller and priced by a bargaining algorithm. The seller's product costs are 50 Coins. Participants receive a list of 11 potential prices $p_{alg} \in \{50, 60, ...150\}$ in steps of 10 and indicate for each whether they are willing to accept it. At the end of the round, their maximum willingness to accept will be compared with the product price. If the product price is higher, then both parties receive 0 Coins. If it is equal or lower than the willingness to accept, then the participant purchases the product for the offered price and earns the value surplus. The bargaining algorithm makes an offer $v \in \{v_0, v_{pd}\}$ with $v_{pd} = v_0 - z > 0$ through price-setting. For example, if the algorithm chooses a price of 100 Coins, then both the participant and the seller earn 50 Coins. If there is no data leak ($1-p$), then $p_{alg} = 100$ and $v = v_0 = 150 - 100 = 50$ Coins. If there is a data leak ($p$), the algorithm bargains more aggressively with $p_{alg} = 100 + 30 = 130$, reflecting that personal information enable personalized pricing that redistributes consumer surplus to seller surplus. In \neutral{}, the algorithm increases prices and hence reduces the offer by $30$ Coins due to the random lottery draw. Note that we do not disclose the outcome of the bargain until after the final willingness-to-pay stage. Thus, we avoid any belief updating regarding the privacy leak.

Finally, in the fourth stage, we elicit subjects' willingness to pay for a privacy label that reveals AI systems with low data security and hence reduces $p$ to $0$ for the second round. This mimics a disclosure institution whereby users can perfectly observe the privacy security of AI products. We use the common Becker-DeGroot-Marschak (BDM) method \citep{becker1964measuring} for an incentive-compatible WTP measurement. Subjects can make a bid $b \in [0, b_{max}]$ where $b_{max} = 2 \times b^* = 30$. Here, $b^* = 15$ is the optimal bid (see Predictions below). Following standard procedure, we set the bidding interval by multiplying the optimal bid by 2. This also allows for ``over-bidding'', as subjects maximum monetary loss from personalization is $\Delta E = 15$ Coins, to account for non-monetary concerns and leave room for non-standard preferences (e.g., regarding risk and ambiguity).  

The bid $b$ is compared to a random draw from the interval $[0, b_{max}]$, and if $b$ is at least as large as the random draw, subjects purchase the privacy label for the price of the random draw. Otherwise, they pay nothing, and receive nothing. To streamline the WTP elicitation, we inform subjects that it is in their best interest to state their true willingness to pay, and display the associated probabilities for each bid to successfully purchase the privacy label. In \neutral{}, we elicit subjects' willingness to pay for the AI recommendation service in the second round without entering the payment lottery. After the bid, we reveal whether the subjects' private data was leaked. In \neutral{}, we reveal whether the subject was randomly drawn during the lottery. Then, we allow all subjects to adjust their prior WTP bid $b$. Finally, we draw the random number, compare it to $b$, and inform subjects about the outcome. This concludes the first round. The second round is largely equivalent to the first one, except for two things. One, subjects learn that they will use a different e-Commerce platform with a different AI system and a different bargaining algorithm in the second stage. They learn that none of these have access to or can utilize their private data. Two, there is no WTP stage, the second round ends with the Bargaining stage and a privacy-leak (lottery outcome) disclosure for subjects who did not purchase the label and chose to use the personalized AI recommendation. The experiment ends with a post-experimental questionnaire that captures ambiguity beliefs via a slider in \ambi{}, betrayal aversion \citep{kormylo2025till,erlei2026betrayal}, and measures privacy concerns using the IUIPC-8 scale \citep{malhotra2004internet,gross2023toward}.

\noindent
\textbf{Technical Implementation.} The experimental platform was developed using a Flask-based web framework with a PostgreSQL database, employing Jinja2 templates to ensure stable cross-platform functionality. Participants first completed a pre-survey administered through the Qualtrics platform before being redirected to the main web application. Experimental integrity was ensured through randomized assignment, which we implemented using the randomizer feature on Qualtrics. The main application received information from the Connect crowdsourcing platform and the Qualtrics Service API, including user preferences and demographic details. This information was then displayed to the participants dynamically, depending on the conditions. Participant progression through the multi-stage experiment was managed via Flask’s session system, which handled treatment randomization, comprehension checks, and decision-making tasks. The platform implemented the Becker-DeGroot-Marschak (BDM) auction mechanism through server-side price generation and incentive-compatible bidding interfaces. Real-time data collection captured all computed economic outcomes, including prices, surplus, and final payments. At the end of web-app-deployed experiment, participants were re-directed to Qualtrics to finish the post-experiment questionnaire.

\subsection{Outcomes of Interest}
Our first metric is the \textbf{share of subjects choosing the AI recommendation} across treatments. Comparing \risk{} with \ambi{} quantifies the impact of the information environment on AI adoption in the presence of unobservable privacy lemon characteristics. Comparing \ppref{} with \psens{} within the same information environment quantifies the effect of privacy data type on AI adoption. Comparing \neutral{} with the privacy conditions disentangles economic and privacy concerns. It allows us to quantify the importance of non-monetary privacy costs for AI adoption in markets with information asymmetries.

The second metric is subjects' \textbf{willingness to pay} for the privacy label. We are interested in two main patterns. First, comparing WTP numbers across treatments gives us cardinal evidence about the importance of different privacy data violations as compared to an economically equivalent, privacy-neutral situation. Second, by comparing subject bids to the theoretical optimal bid (see Predictions below), we can gauge whether people are willing to fully internalize the economic losses from potential privacy leaks.


Third, we can compare the difference $\Delta WTP = WTP_1 - WTP_0$ where $WTP_0$ is the subject's original WTP and $WTP_1$ is the revised WTP after the privacy leak revelation to gauge how strongly people react to actual privacy leaks, conditional on the sensitivity of the data, and benchmark that reaction to a purely economic lottery (\neutral{}). Stronger (weaker) reactions suggest that people underrate (overrate) their predicted dis-utility from privacy leaks.

Finally, we compare \textbf{rejection rates} during the bargaining stage between treatments. In particular, we are interested in subjects' bargaining behavior conditional on the possibility that an automated seller utilizes leaked, private data. Such a context may have substantial consequences for social preferences, which could translate into, for example, more assertive bargaining or a higher propensity to ``punish'' sellers, decreasing overall welfare.

\subsection{Predictions}
In Stage 2, risk-neutral participants maximize their expected payoff conditional on the available information. The expected return is $E(\pi_{s1}) = u_s$ for the standard basket and $E(\pi_{s1,risk}) = u_p - 0.3z - c_i$ or $E(\pi_{s1,ambi}) = u_p - \hat{p}_i z - c_i$ for the personalized basket in \risk{} and \ambi{} respectively. Here, $c_i$ are non-monetary privacy concerns. In \ambi{}, utility depends on subjective probabilities $\hat{p}_i$. If participants maximize their monetary payoff with $c_i = 0$ they always use AI personalization because $E(\pi_{s1,risk}) = 35 - 9 = 26 > u_s = 20$ and $E(\pi_{s1,ambi}) = 35 - 15 = u_s = 20$ for the most pessimistic ambiguity beliefs $\hat{p}_i = 0.5$. Hence, in \risk{}, agents who are only concerned about their economic concerns always use the AI service, and in \ambi{}, the most pessimistic agent is indifferent between using and not-using the AI, while all other types opt for the AI. Non-monetary privacy concerns $c_i$ can alter that relationship if, for instance, $c_i \geq 6$ in \risk{}. Because the literature provides no clear prediction, and behavior depends crucially on (1) mostly unobservable non-monetary concerns and (2) subjective probabilities in \ambi{}, we did not pre-register a directed hypothesis regarding any differences in AI utilization between the two information environments.

In Stage 3, subjects play an ultimatum game against an algorithm framed as a purchasing and price-setting scenario. According to standard game-theoretic predictions, subjects should accept any offer greater than 0, i.e. any price $p_{alg} \leq 140$. Because the prices of the pricing algorithm are always lower than 140 irrespective of $z$, neither the privacy leakage nor the lottery change expected behavior. Subjects always accept the algorithm's pricing offer. However, a long literature in behavioral economics documents that people tend to reject offers that are smaller than 40\% of the whole pie, whereas almost everybody accepts an equal split \citep{erlei2022s}. Furthermore, participants may react more negatively to higher prices induced by a leakage of private data. Hence, if subjects are \textit{behavioral}, rather than pure payoff-maximizers, rejection rates could differ depending on (a) the data leakage and (b) treatments.

In Stage 4, we elicit subjects' willingness to pay for the privacy label. At this point, they do not know whether their data leaked, nor the prices set by the algorithm in Stage 3. Therefore, they only condition their WTP on the expected loss from the leak as communicated in the personalization stage ($\hat{p}_i \times 30$), and do not consider the potential bargaining loss (which would be relevant if they, e.g., accepted a price of 100 but not 130). Intuitively, the label removes the downside of using personalization, by erasing the stochastic monetary loss and $c_i$. During the first personalization stage, subjects weigh these costs against the economic benefits of using AI. This basic trade-off creates a simple revealed-preference insight where participants who chose personalization reveal lower overall privacy dis-utility (lower $c_i$ and/or lower $\hat{p}_i$) because they already preferred personalization. Thus, our main prediction is that \textit{users} of the personalization AI exhibit a strictly lower willingness to pay for the privacy label than \textit{non-users}. Furthermore, non-users should bid exactly $b^* = 15$ to alleviate the dis-utility preventing them from engaging with personalization. Anything more eliminates the benefit of switching from the standard to the personalized basket. Formally, let $\Delta E=15$, $z$ denote the subject's monetary loss per leak in Stage~2, $\hat p_i$ their belief about the leak probability, and $c_i\ge 0$ the non-monetary dis-utility from a privacy leak risk. Round 2 counterfactual payoffs are
\[
E_{\text{no}}=\max\{20,\,35-\hat p_i z -c_i\},\qquad 
E_{\text{yes}}(b)=\max\{20,\,35-b\}.
\]
The indifferent BDM bid satisfies $35-b^*=E_{\text{no}}$, i.e.
\[
b_i^*=35-E_{\text{no}}=\begin{cases}
15, & \text{if } 20\ge 35-\hat p_i z -c_i\quad(\text{no AI w/o label}),\\
\hat p_i z + c_i, & \text{if } 35-\hat p_i z -c_i>20\quad(\text{AI w/o label}),
\end{cases}
\]
with $b_i^*\in[0,15]$. The optimal bid $b^*$ can never be larger than $15$ because, in that case, not using the AI dominates using the AI. Thus non-users bid $\Delta E=15$, whereas users bid their expected leak harm $\hat p_i z + c_i<15$ (which is true through revealed preferences).



\section{Results}
\begin{figure}[h]
    \centering
    \includegraphics[width=\linewidth]{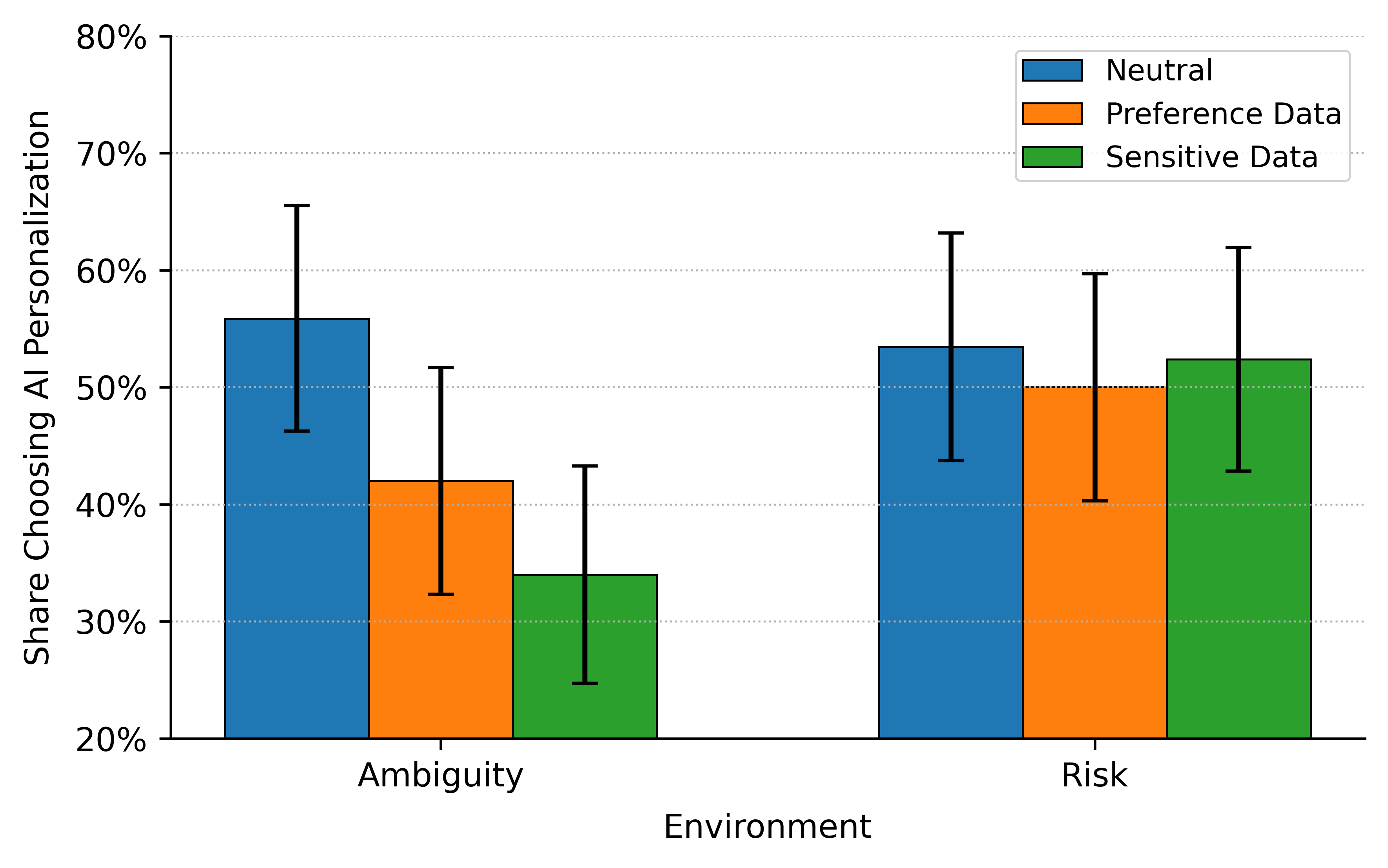}
    \caption{The proportion of subjects consenting to share their data with the AI personalization system.}
    \label{fig:r1_del}
\end{figure}

Figure \ref{fig:r1_del} shows the share of participants relying on the AI personalization system in Stage 2 across the six conditions.

In \risk{}, there are no differences in AI utilization. Across treatments, around 50\% of subjects choose to personalize their basket. Given that the risk-neutral expected payoff of AI personalization is positive (+ 5 Coins), and there are no differences between \neutral{} and the privacy conditions, this points to risk-aversive attitudes among users. In contrast, participants exhibit significant behavioral differences between the \ambi{} conditions ($\, \chi^2 = 10.06, p = 0.007$), with less personalization in both \ppref{} (42\%; $\, \chi^2 = 3.89, p = 0.048$) and \psens{} (34\%; $\, \chi^2 = 9.77, p = 0.002$) than in \neutral{} (56\%). Thus, in information environments where the probability of a leak is not clearly quantifiable, subjects exhibit behavior consistent with non-monetary privacy-related dis-utility. Notably, there are no differences in AI utilization between \neutral{} \risk{} and \neutral{} \ambi{}, suggesting that, in a privacy-neutral setting, neither privacy and ambiguity preferences, nor different information priors under ambiguity, induce behavioral change. Regarding data type, descriptively, participants are more averse to leaks when they submit sensitive demographic data rather than preference data. However, differences in AI utilization are not significant ($\, \chi^2 = 1.36, p = 0.24$), which may be partially related to our statistical power.

\begin{framed}
\begin{itemize}[leftmargin=*]
    \item [] \textbf{Result 1:} In risky information environments with quantifiable probabilities, subjects' AI utilization behavior is not affected by the threat of personal data leaks.
\end{itemize}
\end{framed}

\begin{framed}
\begin{itemize}[leftmargin=*]
    \item [] \textbf{Result 2:} In ambiguous information environments, the threat of a data leak reduces subjects' utilization of AI personalization.
\end{itemize}
\end{framed}

\begin{figure}[t]
    \centering
    \includegraphics[width=0.5\textwidth]{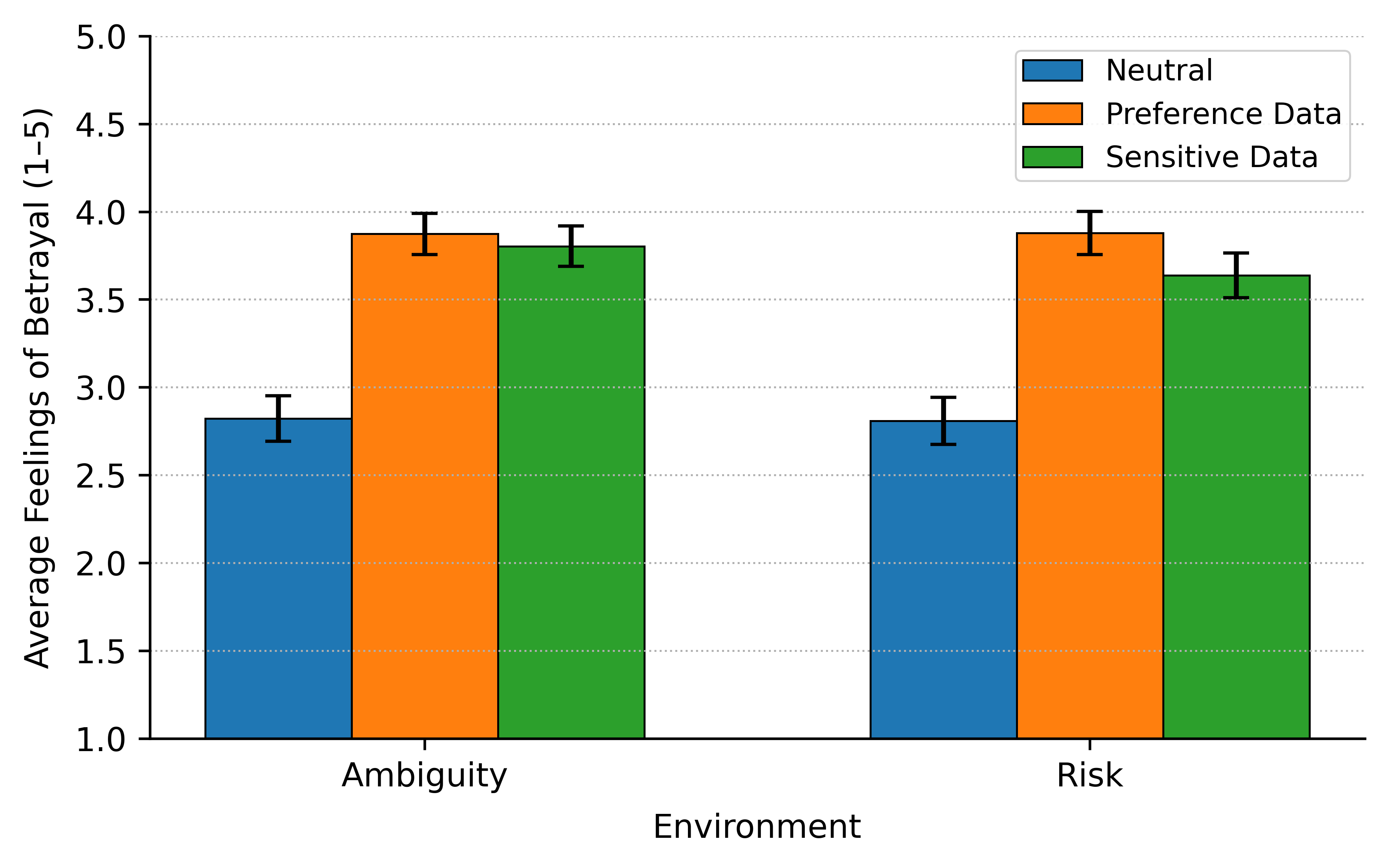}
    \includegraphics[width=0.5\textwidth]{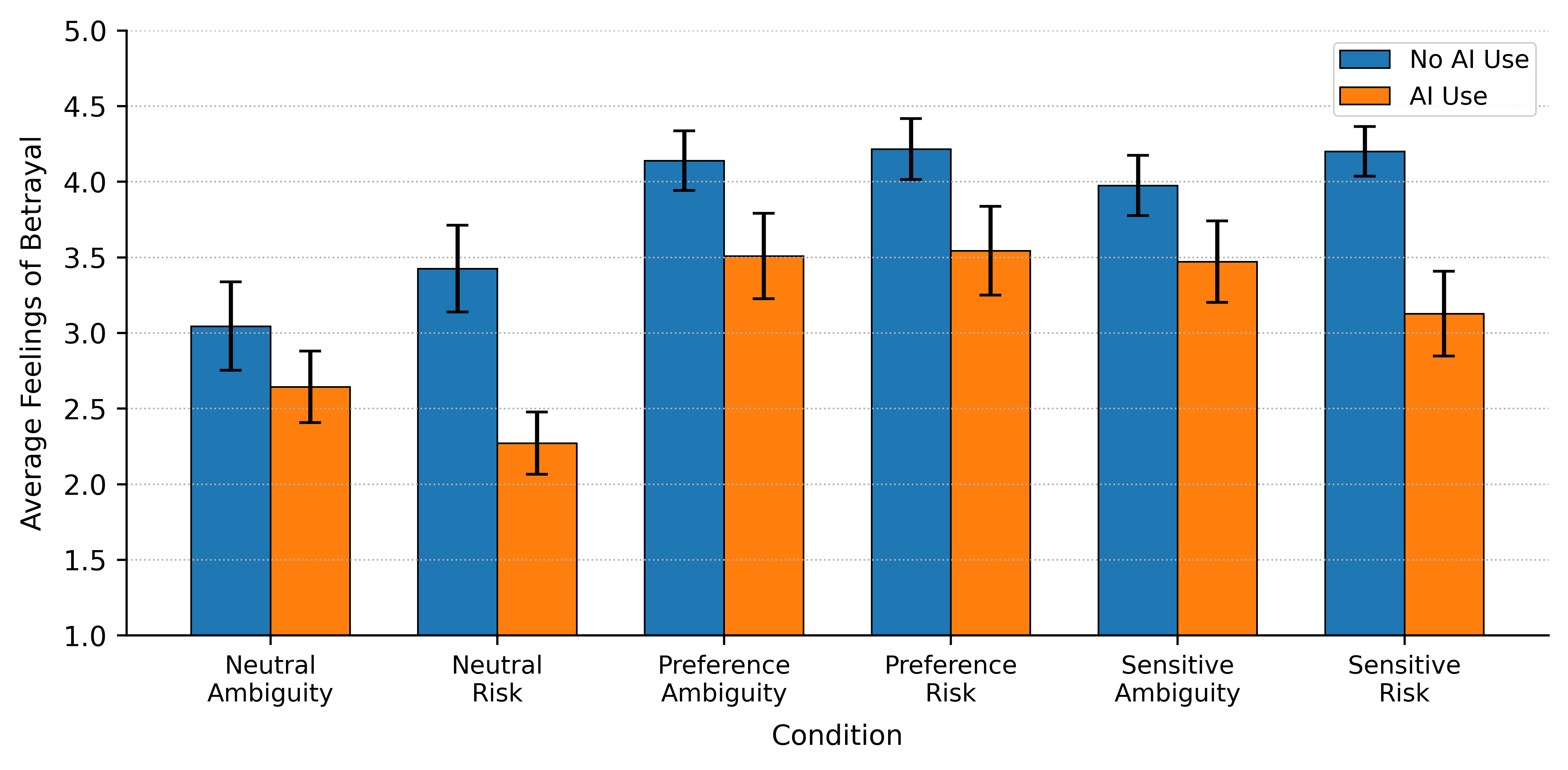}
    \caption{\textbf{Top:} Average feelings of betrayal as measured by the post-experimental questionnaire. Following \citet{kormylo2025till}, we use an average of three questions that elicit feelings of violated trust, feelings of betrayal, and feeling like one made a mistake, after observing the data leak (lottery). \textbf{Bottom:} Average feelings of betrayal across treatments and the AI personalization choice.}
    \label{fig:betrayal}
\end{figure}

To help interpret these results, Figure \ref{fig:betrayal} shows subjects' self-reported feelings of betrayal after a data leak (or lottery draw in \neutral{}). The scores are based on answers from the post-experimental questionnaire using a 5-point Likert-scale. Following \citet{kormylo2025till}, we average three questions about (1) trust violation, (2) feelings of betrayal and (3) feelings of having made a mistake to calculate a composite betrayal score. For each individual question, see Figure \ref{fig:a_betr} in the Appendix. Results show that subjects in the privacy conditions report significantly and substantially stronger betrayal concerns regarding a data leak than subjects in \neutral{} do regarding the lottery draw (see also Table \ref{tab:a_betrayal}). This points to betrayal aversion as a potential source of non-monetary dis-utility \citep{bohnet2004trust}.

Furthermore, individuals who utilize AI personalization report on average significantly lower feelings of betrayal after a potential data leak across all conditions (right panel of Figure \ref{fig:betrayal}). This is consistent with betrayal aversion moderating choice behavior in our experiment. The OLS regression confirms that subjects with higher privacy concerns also report stronger feelings of betrayal, while demographic factors do not appear to play a role. Following expected utility, non-monetary concerns only cause a switch in behavior if they outweigh the perceived economic benefits from AI personalization. Hence, the ambiguous information environment could induce pessimistic subjective probabilities that are skewed towards $\hat{p} = 0.5$, thereby reducing the perceived economic benefits from personalization and allow betrayal aversion to shift AI utilization.

Next, we consider participants' willingness to pay for a privacy label that perfectly identifies ``safe'' AI systems and hence eliminates the threat of a privacy leak. On average, bids $b$ are close to the optimal WTP $b^* = 15$, and there are no relevant treatment differences (see Figure \ref{fig:a1} in the Appendix). This confirms that subjects are generally willing to internalize the negative effects of unsafe AI systems by financing costly transparency institutions such as labels. Pooling WTP data across conditions shows that, in contrast to our predictions, users (17.7) are willing to pay more on average than non-users (13.6, $\, t = 5.85, p < 0.001$). In fact, users are willing to pay more than the maximal return (15 - $c_i$), indicating that users \textit{over-bid} for a privacy label that eliminates uncertainty about privacy concerns. This patterns holds for both \risk{} (user: 17.3, non-user: 13.2) and \ambi{} (user: 18.3, non-user: 14.04). After learning that a leak occurred, participants on average increase $b$ by 1.9, with substantial heterogeneity between the neutral and privacy conditions (\neutral{:} -0.23, \ppref{:} 3.65 ($t = 2.01, p = 0.05$), \psens{:} 2.86 ($t =2.15, p = 0.03$)). If the leak did not occur, participants reduce their bid by, on average, 1.07, with no treatment differences (\neutral{:} -1.15, \ppref{:} -0.93, \psens{:} -1.07). This, again, points to non-monetary aversions against privacy leaks.

\begin{figure}[t]
    \centering
    \includegraphics[width=0.45\textwidth]{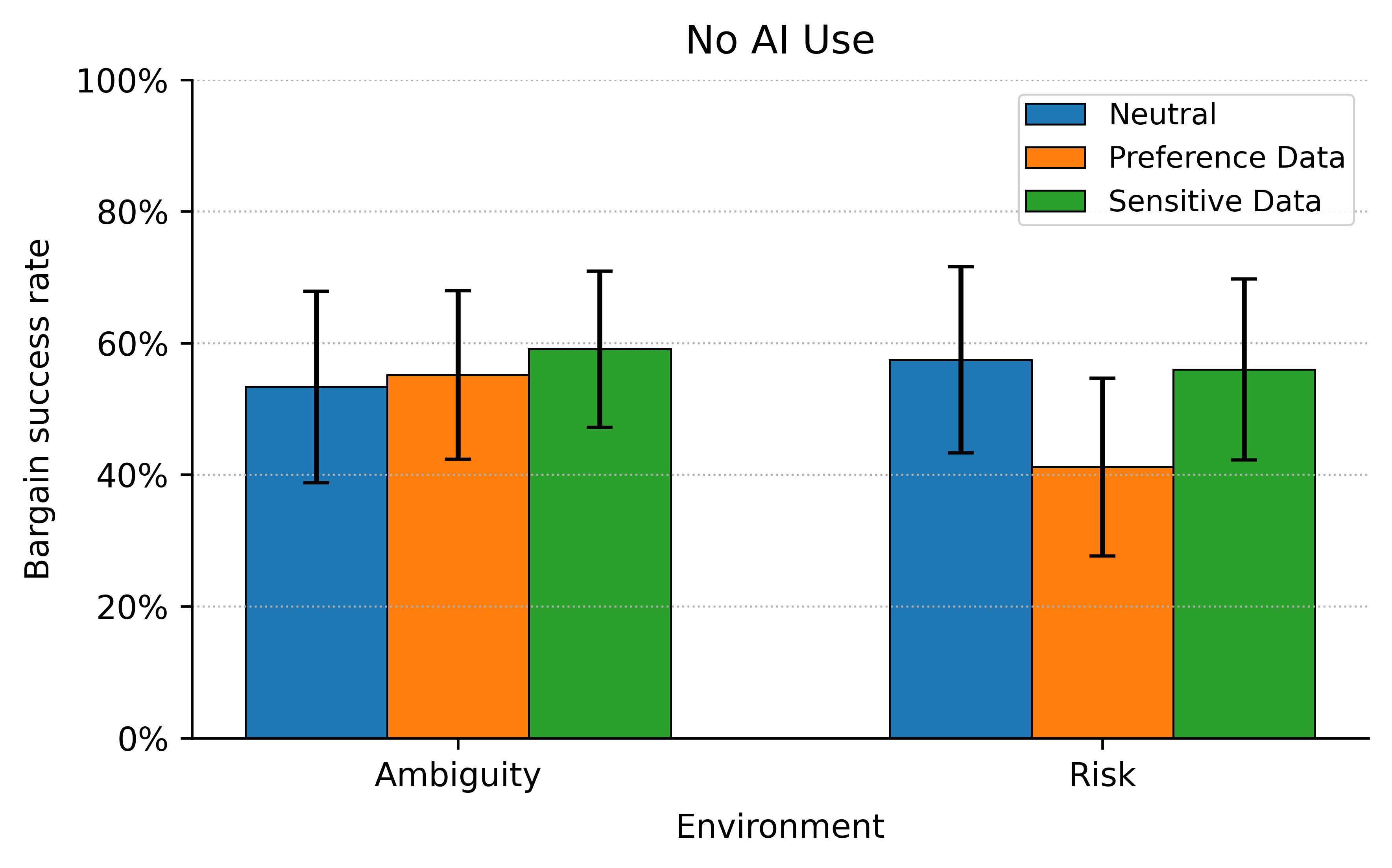}
    \includegraphics[width=0.45\textwidth]{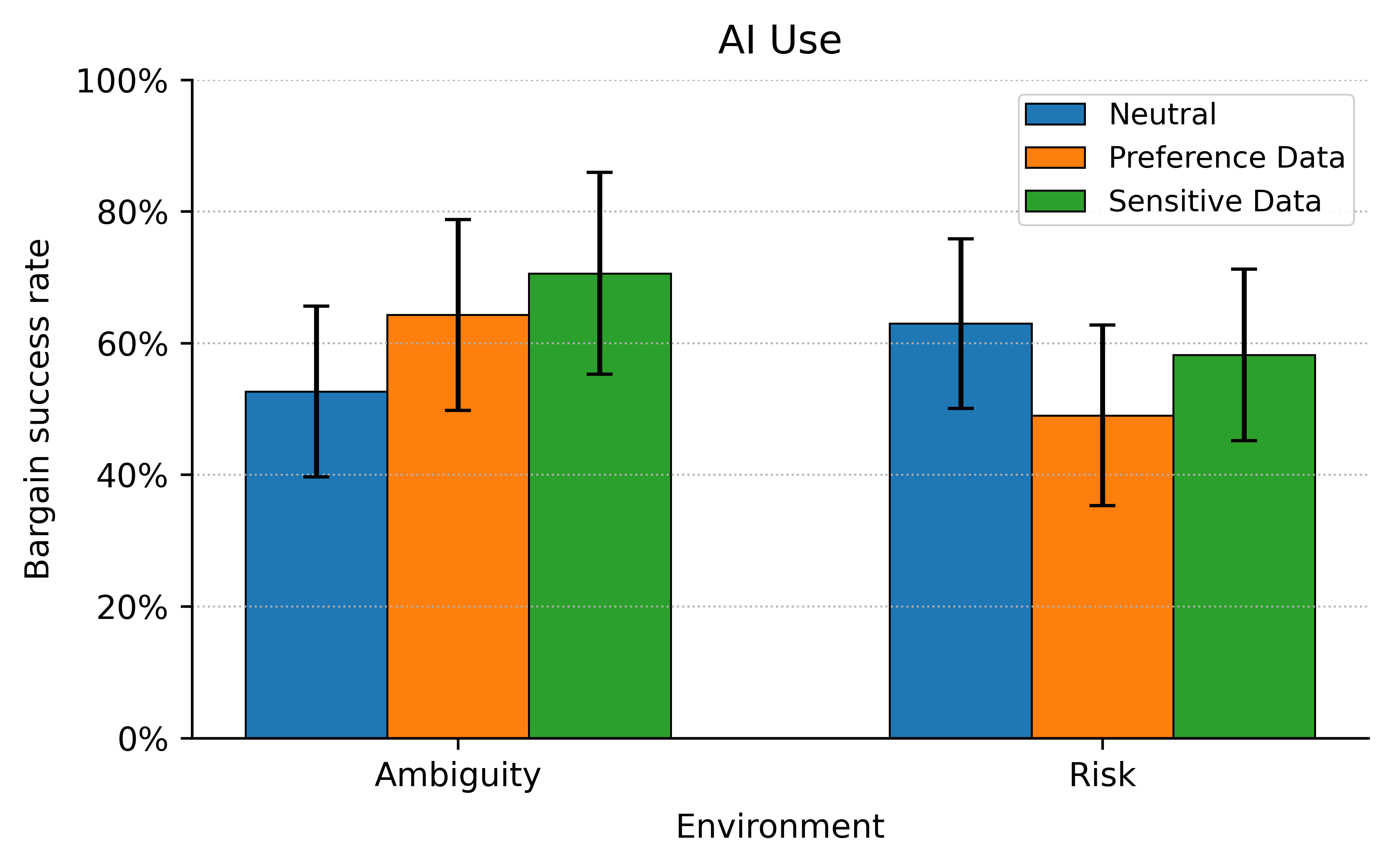}
    \caption{Rejection rates in the framed ultimatum game.}
    \label{fig:rejection_rates}
\end{figure}

\begin{framed}
\begin{itemize}[leftmargin=*]
    \item [] \textbf{Result 3:} On average, participants demonstrate a willingness to absorb the economic costs of mitigating privacy risks by financially supporting transparency mechanisms, such as third-party verification, that ensure secure handling of personal data by providers. 
\end{itemize}    
\end{framed}

\begin{framed}
\begin{itemize}[leftmargin=*]
    \item [] \textbf{Result 4:} Participants exhibit a strong preference for certainty in privacy protection, willingly over-paying for privacy labels that fully eliminate ambiguity about data leak risks. Their willingness to pay increases in response to actual privacy threats rather than abstract probabilistic scenarios (i.e., a lottery draw) highlighting the importance of clear, actionable risk communication in interface design. 
\end{itemize}
\end{framed}

Third, we look at bargaining behavior and rejection rates (Figure \ref{fig:rejection_rates} and see Figure \ref{fig:a2} in the Appendix for subjects' average maximally accepted prices which track rejection rates).
Overall, rejection rates are comparable across conditions, and there is little evidence that the threat of data leaks affects bargaining behavior. Subjects who chose to rely on AI personalization do, on average, accept higher prices in \ambi{} but not in \risk{} (Table \ref{tab:a1}). This does not depend on the privacy condition, confirming that ambiguity and informational concerns drive less aggressive user bargaining, rather than privacy concerns. Consequently, there are no differences in bargaining outcomes (Table \ref{fig:a2}), although qualitatively, less aggressive user bargaining after AI personalization correlates with descriptively more favorable bargaining outcomes in the privacy conditions under \ambi{} (53\% vs. 64\% vs. 71\%). 


\begin{framed}
\begin{itemize}[leftmargin=*]
    \item [] \textbf{Result 5:} When users face ambiguity about potential price increases during bargaining, they lower their minimum acceptable price, suggesting that uncertainty shapes negotiation strategies. However, the mere threat of a data leak does not alter bargaining behavior. 
\end{itemize}    
\end{framed}

Table \ref{tab:r2_del} shows results from a logit regression on the user choice to use AI personalization in round 2. 

\begin{table}[!t]
\centering
\caption{Logit regression of AI use in round 2}
\label{tab:r2_del}
\begin{tabular}{l*{3}{c}}
\toprule
 & \textbf{Ambiguity} & \textbf{Risk} & \textbf{Pooled} \\
\midrule
Preference\_Data   & 2.039* & 1.124  & 1.524 \\
                        & (0.635) & (0.351) & (0.334) \\
Sensitive\_Data     & 1.646   & 1.541  & 1.609* \\
                        & (0.513) & (0.488) & (0.356) \\
Data\_Leak       & 4.750*** & 8.689*** & 6.022*** \\
                        & (2.327)  & (5.499)  & (2.301) \\
Privacy\_Label      & 3.309*** & 4.302*** & 3.763*** \\
                        & (0.845)  & (1.130)  & (0.687) \\
\midrule
Observations            & 302      & 308      & 610 \\
\bottomrule
\multicolumn{4}{p{0.95\linewidth}}{\footnotesize Odds ratios; standard errors are delta-method SEs on the OR scale. Baseline framing: \neutral{}.}\\
\multicolumn{4}{p{0.95\linewidth}}{\footnotesize * $p<0.05$, **$p<0.01$, ***$p < 0.001$.}
\end{tabular}
\end{table}

There are two main results. One, compared to round 1, the negative effect of being exposed to a data leak on AI personalization vanishes, and, if anything, reverses. Two, having been exposed to a data leak in round 1 strongly predicts using the AI personalization in round 2. Thus, despite being informed that round 2 involves new pricing algorithms from different third-party sellers, experiencing a data leak does not appear to deter users from using AI personalization, and likely even increases AI use. This may speak to priorly documented habituation or resignation patterns, where people become less sensitive to data leaks after experiencing them (see e.g. \citet{chen2023research,turjeman2024data}). As expected, subjects who purchased the privacy label also relied more on AI personalization in round 2 (80\% vs. 43\%). 


\section{Discussion}
There has been little empirical research about how the threat of personal data leaks affects user behavior in the context of AI systems. In addition, prior work has not controlled for or manipulated the specific informational disposition of the people potentially affected by such a leak. This paper, therefore, uses an experimental approach to gauge how people react to potential privacy leaks in the context of AI personalization. We disentangle privacy from economic concerns by comparing a data-leak frame with a neutral lottery gamble frame. Furthermore, we endow participants with information about the likelihood of a leak, inducing either a risky or an ambiguous choice environment. Our abstraction captures a decision dilemma that is ubiquitous in the context of personalization \citep{asthana2024know}. People with different private signals and different levels of expertise must trade-off a certain gain through the disclosure of personal information against the possibility that their data may later be observed and used by other agents. 

\begin{quote}
    \textit{\textbf{RQ1}: How does the threat of a personal data leak affect AI adoption under risk and ambiguity?}
\end{quote}

In our experiment, being stochastically exposed to data leaks only negatively affects AI-personalization adoption if subjects cannot ascribe definite probabilities to the leak outcome. People who are fully informed about the risks do not differentiate between the neutral and the data frame. Furthermore, the negative effect of a potential data leak is not confined to sensitive demographic data, but also occurs when subjects must share anonymized preference data. Descriptively, the effect is stronger for identity-related data. To our knowledge, this is the first empirical study that carefully disentangles both the information environment and the threatened data type. The fact that privacy behavior systematically depends on the information endowment, where the threat of a personal data leak only reduces AI utilization if subjects are uncertain about the true underlying probabilities, opens up many novel questions about the stability and endogeneity of privacy preferences. Note that, because AI personalization use is contingent on the data leak threat in \ambi{}, and because the degree of ambiguity is the same between treatments, classic ambiguity-aversion cannot explain these patterns. This is further supported by the fact that there is no difference in AI utilization between \neutral{} \ambi{} and \neutral{} \risk{}. One potential interpretation is that uncertainty about the AI system's underlying leak probability induces pessimistic subjective probabilities \citep{einhorn1985ambiguity,bier1994ambiguity}. There is empirical evidence that ambiguity-averse agents form downward-biased expectations (i.e., pessimism) \citep{zhang2022subjective,pulford2009short,ilut2022modeling}, particularly if they compare the ambiguous option with a less ambiguous one \citep{fox1995ambiguity}. If true, then subjects in \ambi{} could react substantially stronger to a potential data threat because they underestimate the economic gains from the ``personalization lottery''. For instance, in \risk{}, $c_i \geq 6$ must hold for subjects to forego AI personalization. A lower subjective probability decreases the expected return and hence the critical value of $c_i$ such that the anticipated non-monetary disutility from a leak changes the agent's optimal choice. In the most pessimistic case, $\hat{p}_i = 0.5$, the agent is indifferent between using and not using the AI. Here, any non-monetary concerns shift the optimal choice towards the standard basket without AI. This is consistent with results from the post-experimental questionnaire, where subjects in the privacy treatments generally indicated substantially stronger concerns about being betrayed and having their trust violated. While these do not differ between \risk{} and \ambi{}, they indicate a kind of non-monetary disutility which is only present when subjects face a data leak threat. As shown above, this only affects choice behavior if the expected economic gains from AI personalization are low enough, which is true for downward-biased subjective probabilities in \ambi{}. Note that under betrayal-aversion, reactions may be much stronger if personal data leaks to human agents rather than a pricing algorithm. While there has been very little work on the interaction of betrayal aversion and AI or algorithms, two recent papers suggest that in the context of financial advice and trust-giving, using an algorithmic advisor or an LLM agent strongly alleviates betrayal aversion \citep{kormylo2025till,erlei2026betrayal}.

\vspace{.5em}
\begin{quote}
\textit{\textbf{RQ2}: How does the threat of a personal data leak with subsequent economic consequences affect Human-Algorithm bargaining?}   
\end{quote}

Sellers increasingly rely on personal data to set personalized prices or target advertisements. In that context, knowing how the possibility of a data leak with subsequent strategic (economic) consequences affects user behavior is a consequential question. Our results suggest that people do not socially retaliate against pricing algorithms that possibly use private data to set their prices. If anything, people who believe that prices may be adjusted upwards due to the algorithm's access to private data can be willing to accept a \textit{lower} profit share, i.e., willing to pay a higher price. This provides the first preliminary evidence that personalized bargaining strategies are not negatively affected by consumer privacy behavior.

\vspace{.5em}
\begin{quote}
\textit{\textbf{RQ3}: How much are users and non-users willing to pay for a privacy disclosure label that reveals whether an AI system handles personal data safely?}
\end{quote}

Theory predicts that users should be willing to pay less for a disclosure label than non-users. Our experiment finds the opposite. On average, users are willing to over-pay for the privacy label, suggesting potentially strong incentives for providers and sellers to invest in costly data security. Furthermore, across the whole sample, users are generally willing to completely internalize all negative economic costs associated with data leaks. The average label bid is close to the theoretically optimal bid. Hence, in our experiment and contrary to some prior evidence, people are willing to pay for data security, especially those who exhibit AI personalization preferences. One potential interpretation of the data is that preferences for or against personalization are partially a character trait. Then, people with a taste for personalization may simultaneously be more likely to accept the threat of a data leak and be willing to pay more for a perfect disclosure institution, despite their revealed preferences suggesting lower privacy concerns. Regarding previous studies suggesting a comparatively low willingness to pay for privacy protection, results from our experiment may be driven by a few distinct factors. One, the analyzed privacy label removes uncertainty by guaranteeing that no leak will occur. This may be a particularly strong signal that cannot be replicated by many ``real-world'' verification labels. Second, the consequences in our study are immediate and material. In reality, any potential leak is unlikely to occur immediately, leaving ample room for time discounting \citep{frederick2002time}. Third, the consequences in this paper are relatively simple to understand and calculate, whereas in reality, privacy breaches and leaks exhibit a plethora of potential outcomes that can occur at multiple levels \citep{acquisti2005privacy}. This impedes any economic calculation.

\subsection{Implications}
First, our results show that ambiguity about privacy risk, rather than risk per se, affects AI adoption. From an HCI perspective, this argues for the design of interfaces that reduce user ambiguity by, e.g., presenting concrete probability information (where available) or confidence intervals, rather than convenient but ``generic'' warnings. To support informed AI adoption under privacy uncertainty, interfaces should prioritize transparency and externalizing trustworthiness cues~\cite{zieglmeier2021designing,mehrotra2024systematic}. Designers can potentially use graded risk indicators and confidence visualizations to reduce ambiguity, integrate verified privacy badges as clear trust signals, and employ progressive disclosure for detailed explanations of the information environments without overwhelming users~\cite{springer2020progressive}. Cost trade-offs for privacy verification should be made explicit through interactive comparisons, and scenario previews can help users understand the impact of data-sharing choices on personalization and risk.

Second, the willingness-to-pay results indicate that many users will internalize privacy externalities when offered credible, verifiable assurance. This supports integrating third-party verification directly into critical flows. Following our information treatments, these decisions may be more convincing if communicated as an insurance against ambiguity.

Third, privacy-related user experience decisions may anticipate the downstream effects of reduced user bargaining assertiveness under ambiguity. Possible solutions include the early elicitation of reservation prices to protect users from the negative effects of ambiguity or post-bargain feedback that relieves ambiguity.

Fourth, our effects hold for both data types, with modest descriptive differences. This speaks to potentially broadening the scope of privacy-preserving mechanisms (e.g., defaults or consents) that go beyond personal identity or demographic data and give users agency over their full choice data.

From a theoretical point of view, our findings refine the privacy–personalization literature by separating the ``dis-utility from sharing'' from the ``dis-utility from potential leakage'', showing that the latter is strongly moderated by ambiguity. Hence, any privacy calculus designed to explain or predict user behavior \citep{knijnenburg2013preference,zhang2016privacy,dinev2006extended} may benefit from an explicit ``ambiguity premium'' where choices depend on beliefs and belief precision. Furthermore, our documented WTP patterns suggest that people pay for the resolution of uncertainty, which is why disclosure institutions such as labels and third-party audits should be analyzed and interpreted as information goods \citep{varian2000versioning}, rather than objective artifacts.

Finally, from a policy perspective, our findings suggest that regulatory frameworks should incentivize transparency standards and verification mechanisms, such as third-party privacy certifications, to reduce ambiguity and foster trust. Policies could also encourage cost-sharing models for privacy verification, ensuring that providers are not disproportionately burdened while enabling consumers to make informed choices. Additionally, guidelines for clear communication of uncertainty in digital interfaces could become part of privacy compliance standards. 

Collectively, our findings in fact reinforce the objectives of the EU Digital Services Act (DSA), which emphasizes transparency, accountability, and user empowerment in digital services~\cite{turillazzi2023digital}. By showing that ambiguity about privacy risks significantly reduces adoption of AI personalization, we highlight the need for regulatory standards that mandate clear, user-facing disclosures of data practices and risk levels~\cite{sekwenz2025digital}. Verified privacy certifications, akin to trust badges, should be incentivized under the DSA to reduce uncertainty and foster trust, while cost-sharing models can ensure that compliance does not disproportionately burden providers of personalized AI systems. Incorporating methods for communicating uncertainty into DSA compliance would align with its user-centric design principles, enabling interfaces that support informed decision-making.

\subsection{Caveats, Limitations, and Future Work}
\textbf{Artificial Data Leak.} One primary limitation of this paper is the way our data leak is operationalized. Because we cannot ``really'' leak participants' private information to other humans or outside entities, we chose to introduce price-setting algorithms that observe and strategically utilize personal data. This, however, is an abstraction of real-world data leaks in which a variety of (potentially bad-faith) actors may gain access to user data. This includes algorithmic and AI tools, but also other humans. It is likely that user reactions will be stronger if their leaked data can be observed by other humans, or if other humans strategically use their leaked data to profit from it. Furthermore, in reality, the consequences of a data leak are usually much less deterministic and predictable. Consumers do not know who will, at some point, gain access to their data, or what they may do with it. Therefore, our experiment only captures a specific and relatively narrow type of behavioral privacy-trade-off. 

\noindent
\textbf{Sample.} Our sample primarily includes participants whose first language is English. This is a notable limitation because privacy attitudes and risk perceptions are shaped by cultural norms and linguistic framing \citep{harris2003privacy}. Future research should examine cross-cultural differences in privacy-related decision-making and explore how interface design can adapt to diverse cultural contexts to ensure the broader applicability of transparency and trust mechanisms.

\noindent
\textbf{Complexity.} Our experiment deliberately incorporates forward-looking decision problems that require participants to consider not only potential price increases but also how those prices influence their bargaining success. This complexity reflects real-world negotiation dynamics and enhances external validity, offering richer insights into strategic behavior under privacy risk. A more streamlined design that controls for participants' strategic behavior by simply increasing prices with a clear observable income effect may produce stronger causal estimates at the cost of external validity. 

\noindent
\textbf{Statistical Power.}
Our study was adequately powered to detect main effects, but 
some null effects (e.g., \ppref{} vs. \psens{}) may be underpowered for moderate interactions with \risk{}/\ambi{} or heterogeneous treatment effects by baseline privacy concerns. While this is a common challenge in complex behavioral experiments, our results provide a solid foundation for future work. Pre-registered, higher-powered designs with planned interaction tests and hierarchical modeling could sharpen these estimates and uncover more nuanced patterns.

\noindent
\textbf{Perfect verification assumption.}
Our study models the ``privacy label'' as perfectly informative and fully trusted, which simplifies interpretation and isolates behavioral responses to certainty. While real-world labels often face compliance gaps and credibility challenges, this assumption allows us to establish an upper bound on the potential effectiveness of verified disclosures. Future work should examine how varying levels of trust and imperfect verification influence user behavior, and explore design strategies that can mitigate these real-world limitations.

\noindent
\textbf{Data Sharing in \neutral{}.} Our experiment aims to isolate the effect of data leak threats on AI personalization use if people must share their private data. Therefore, participants in \neutral{} also always share their data. Here, we chose \psens{} over \ppref{} to avoid inflating our results, as we anticipated stronger user aversions towards sharing sensitive data. This design choice enhances internal validity, and our results descriptively confirm that. Thus, if data sharing itself negatively affects AI use, and people are more hesitant to use their demographic data, we may be underestimating the difference between \neutral{} and \ppref{}. In the imminent future, we aim to disentangle the baseline reluctance of users to share data from reactions to privacy threats, and investigate adaptive interfaces that accommodate varying comfort levels with different data types.

\section{Conclusions}
This paper demonstrates that the behavioral effects of privacy leak threats depend critically on the information environment, rather than privacy preferences alone. While participants exhibit no detectable privacy-related disutility under risk, ambiguity about leak probabilities substantially reduces AI adoption across both preference and sensitive data types. Users systematically over-pay for privacy disclosure labels, suggesting that market-based solutions through credible third-party verification may be economically viable. Hence, policy interventions should focus on reducing user ambiguity through standardized probability disclosure rather than simply restricting data collection practices. For designers and policymakers alike, our results highlight the importance of transparency and uncertainty communication in fostering informed and appropriate adoption of AI systems.

\begin{acks}
We thank all the anonymous participants in our study. This work was partially supported by the TU Delft AI Initiative, the Model Driven Decisions Lab (\textit{MoDDL}), the Robust LTP GENIUS Lab, and the \textit{ProtectMe} Convergence Flagship. 
\end{acks}

\bibliographystyle{ACM-Reference-Format}
\bibliography{sample-base}

\clearpage
\section*{Appendix}

\begin{figure}[!htbp]
    \centering
    \includegraphics[width=0.45\textwidth]{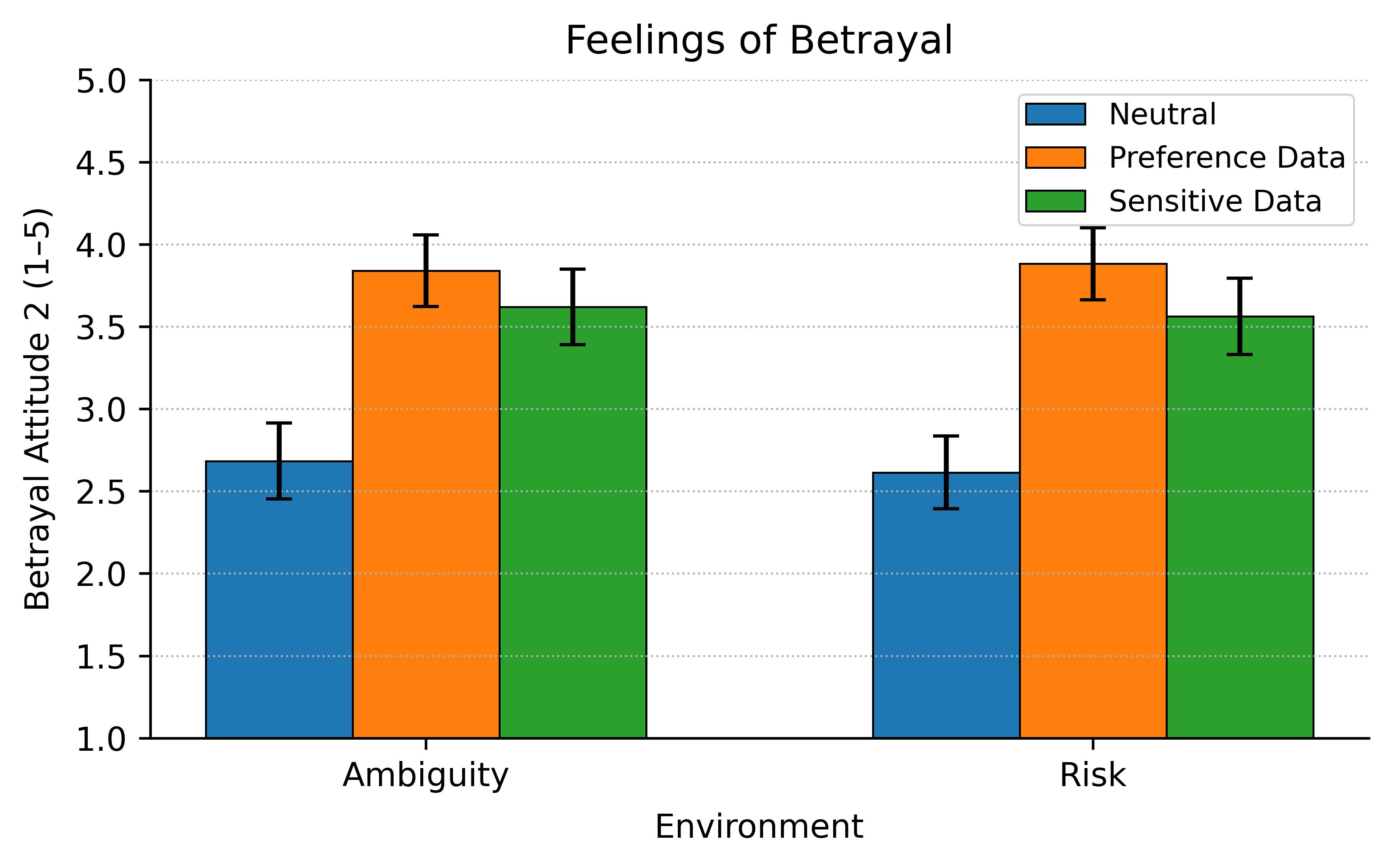}
    \includegraphics[width=0.45\textwidth]{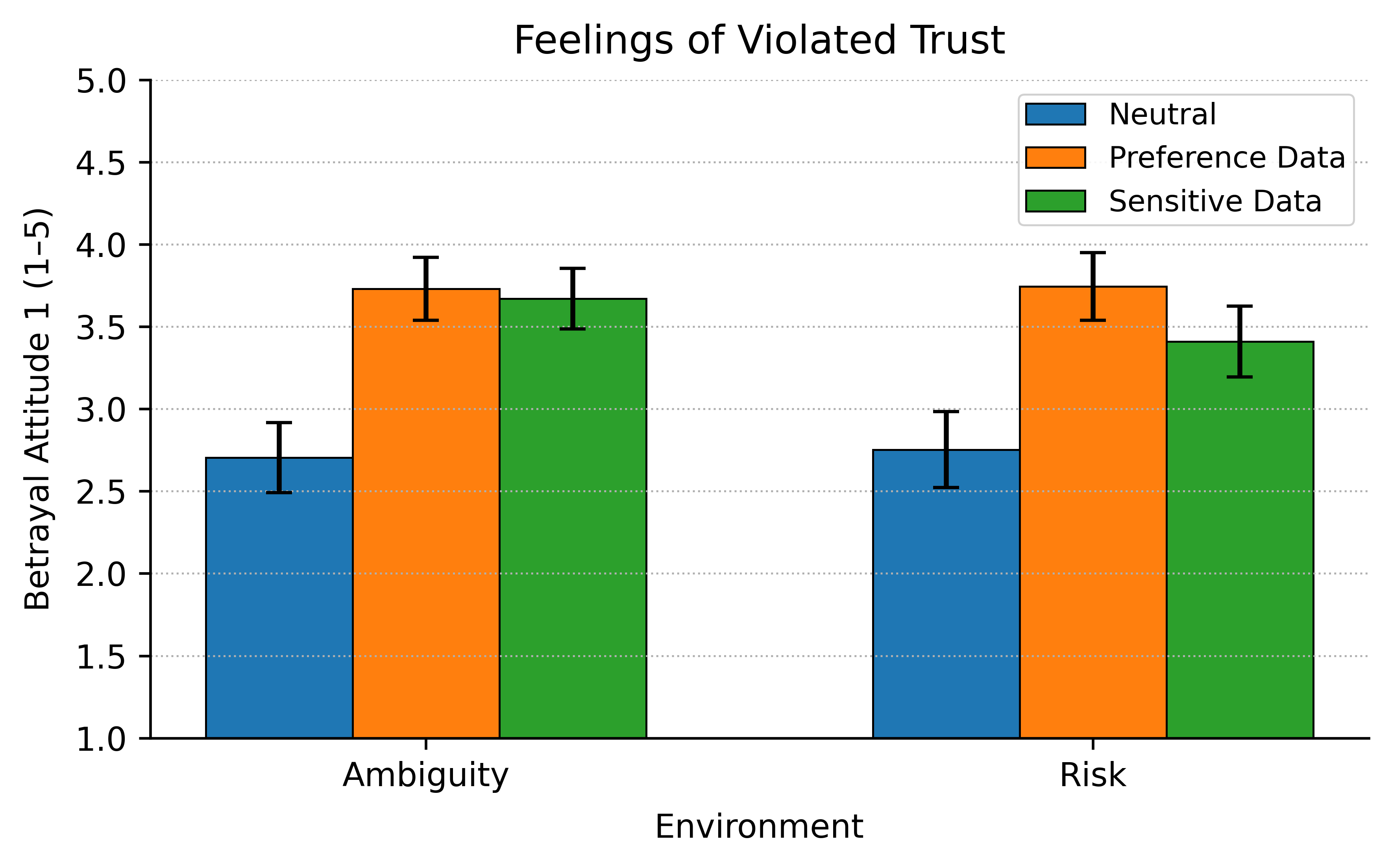}
    \includegraphics[width=0.45\textwidth]{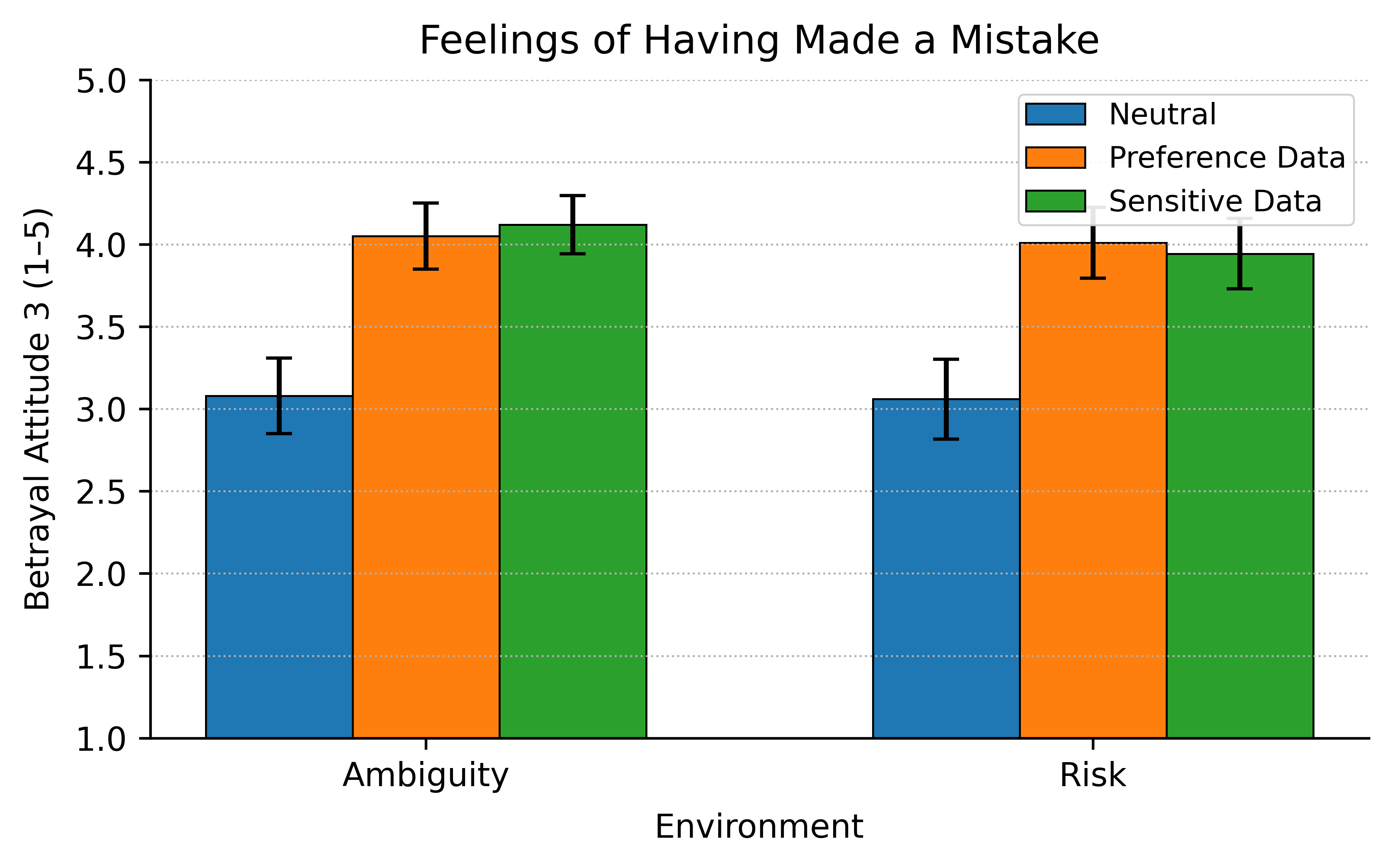}    
    \caption{Feelings of (1) betrayal, (2) trust violations and (3) having made a mistake (from left to right) according to self-reported measures in the post-experiment questionnaire.}
    \label{fig:a_betr}
\end{figure}

\begin{table}[!ht]
\centering
\caption{OLS regression on subjects' self-reported feelings of betrayal.}
\label{tab:a_betrayal}
\begin{tabular}{lccc}
\toprule
 & \textbf{Betrayal} & \textbf{Betrayal} & \textbf{Betrayal} \\
 & \textbf{(1)} & \textbf{(2)} & \textbf{(3)} \\
\midrule
Preference Data              & 1.061***    & 0.958***    & 0.957*** \\
                             & (0.097)     & (0.089)     & (0.092)  \\
Sensitive Data               & 0.904***    & 0.835***    & 0.834*** \\
                             & (0.097)     & (0.086)     & (0.089)  \\
Risk Treatment               & $-$0.058    & $-$0.014    & $-$0.028 \\
                             & (0.078)     & (0.068)     & (0.070)  \\
Used AI                      &             & $-$0.691*** & $-$0.708*** \\
                             &             & (0.071)     & (0.074)  \\
Privacy Attitudes            &             & 0.432***    & 0.434*** \\
                             &             & (0.059)     & (0.062)  \\
Risk Preferences             &             &             & 0.034   \\
                             &             &             & (0.018)  \\      
Time Discounting $\beta$     &             &             & $-$0.330 \\
                             &             &             & (0.256)  \\                               
Constant                     & 2.844***    & 0.622*      & 0.292    \\
                             & (0.079)     & (0.357)     & (0.651)  \\
\midrule
Demographic Controls         & No          & No          & Yes      \\                            
Observations                 & 609         & 609         & 592      \\
R-squared                    & 0.192       & 0.383       &  0.403        \\
\bottomrule
\multicolumn{4}{l}{\footnotesize Cluster-robust standard errors in parentheses. Baseline framing: \neutral{}.} \\
\multicolumn{4}{l}{\footnotesize The dependent variable is the composite betrayal aversion score calculated by} \\
\multicolumn{4}{l}{\footnotesize using the average of our three betrayal questions.}\\
\multicolumn{4}{l}{\footnotesize * $p<0.05$, ** $p<0.01$, *** $p<0.001$.}
\end{tabular}
\end{table}

\begin{figure}[!htbp]
    \centering
    \includegraphics[width=\linewidth]{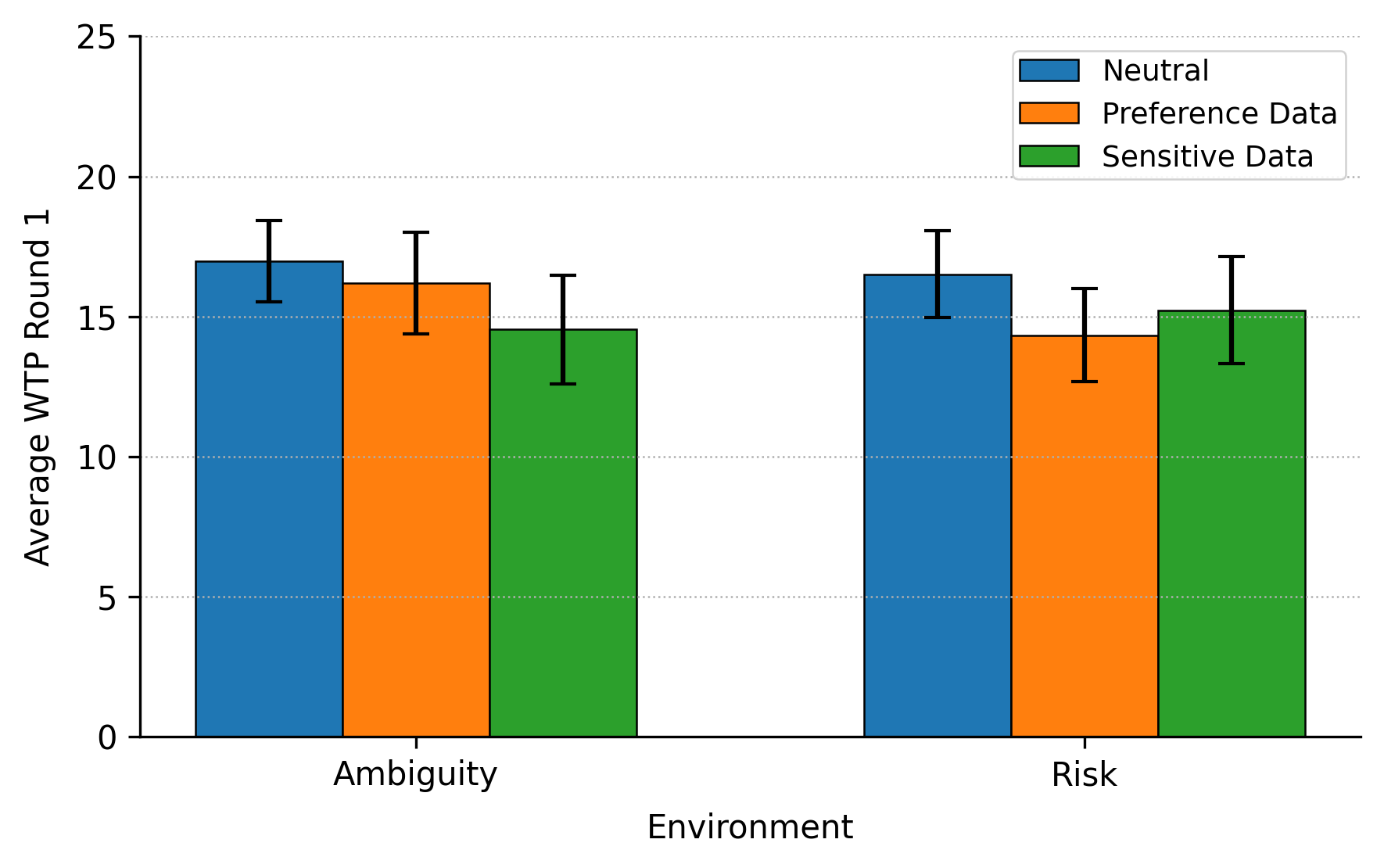}
    \caption{Average WTP for the privacy label in round 1.}
    \label{fig:a1}
\end{figure}

\begin{figure}[!htbp]
    \centering
    \includegraphics[width=0.45\textwidth]{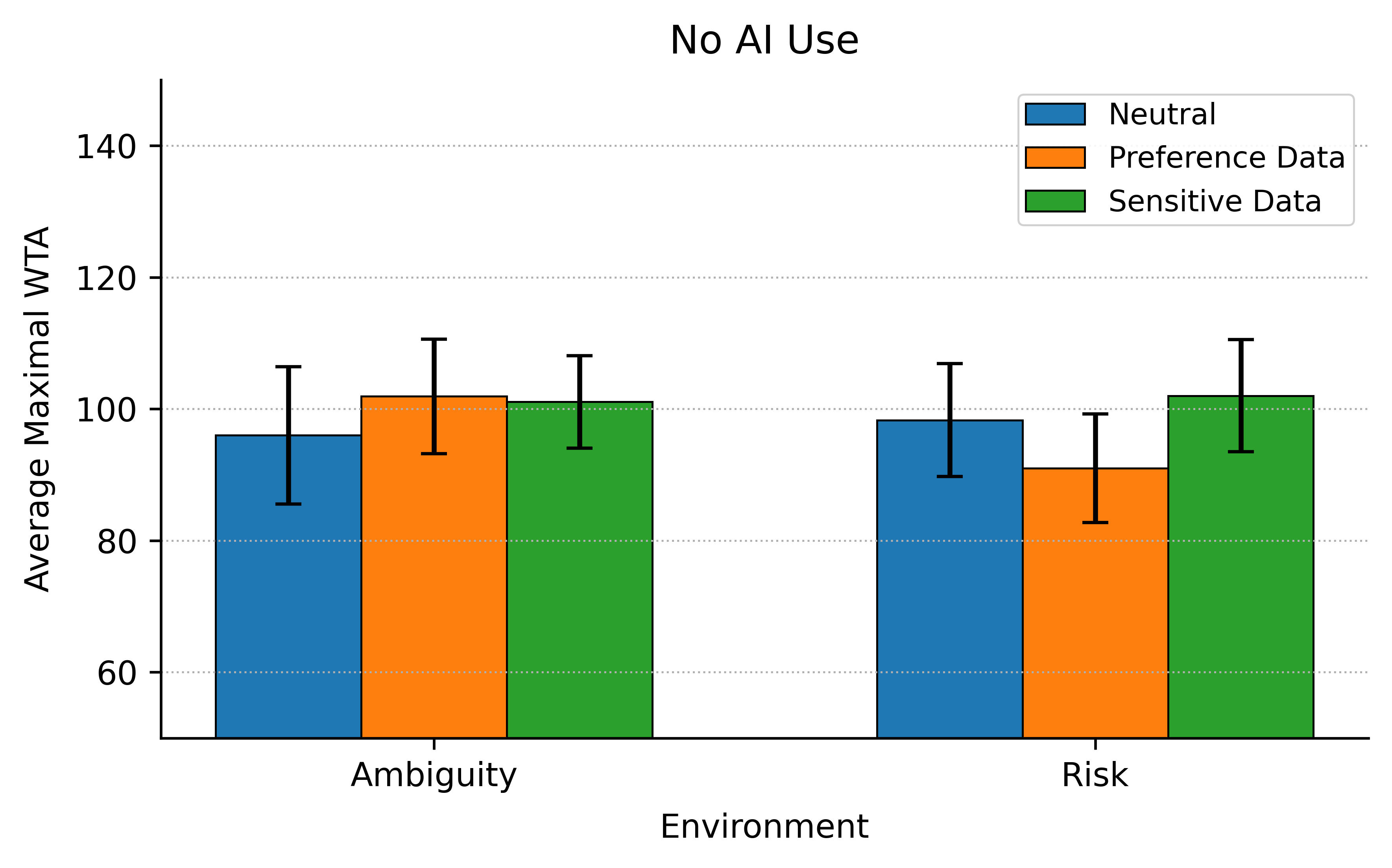}
    \includegraphics[width=0.45\textwidth]{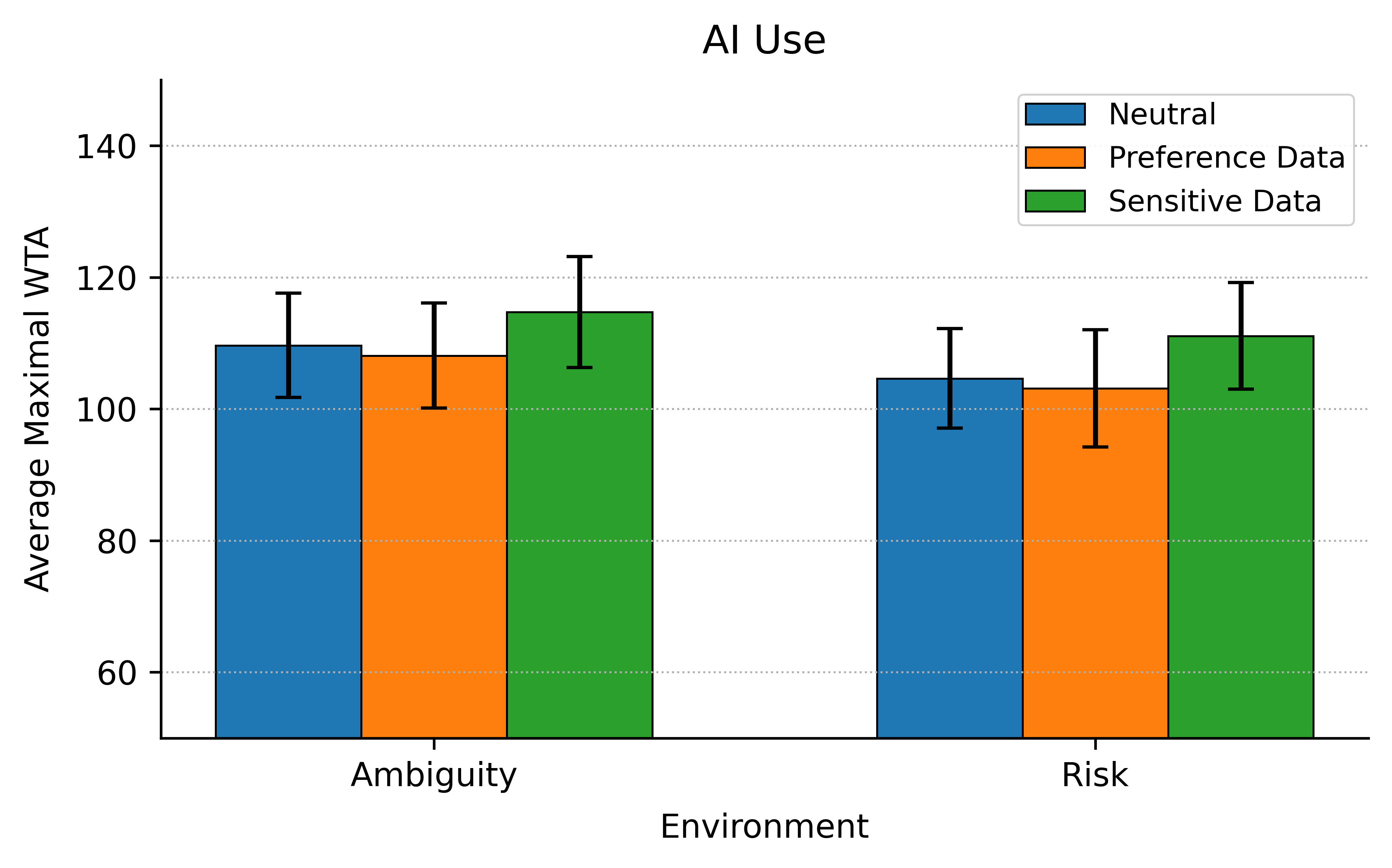}
    \caption{Average WTA for the voucher.}
    \label{fig:a2}
\end{figure}

\begin{table}[!ht]
\centering
\caption{OLS regression on subjects' willingness to accept during the price bargain.}
\label{tab:a1}
\begin{tabular}{lcc}
\toprule
 & \textbf{Risk} & \textbf{Ambiguity} \\
\midrule
Used\_AI                  & 6.33        & 13.65** \\
                             & (5.83)      & (6.68)  \\
Preference Data        & -7.32       & 5.90    \\
                             & (6.07)      & (6.92)  \\
Sensitive Data          & 3.70        & 5.06    \\
                             & (6.16)      & (6.42)  \\
Used\_AI $\times$ Preference Data & 5.83        & -7.45   \\
                             & (8.51)      & (8.98)  \\
Used\_AI $\times$ Sensitive Data   & 2.76        & 0.00    \\
                             & (8.37)      & (8.71)  \\
Constant                     & 98.30***    & 96.00*** \\
                             & (4.36)      & (5.32)  \\
\midrule
Observations                 & 308         & 302      \\
\bottomrule
\multicolumn{3}{l}{\footnotesize Cluster-robust standard errors in parentheses. Baseline framing: \neutral{}.}\\
\multicolumn{3}{l}{\footnotesize * $p<0.10$, ** $p<0.05$, *** $p<0.01$.}
\end{tabular}
\end{table}

\begin{table}[!ht]
\centering
\caption{Logit models of successful bargains (odds ratios)}
\label{tab:a2}
\footnotesize
\setlength{\tabcolsep}{4pt} 
\begin{tabular}{lcccc}
\toprule
 & \textbf{Ambiguity} & \textbf{Ambiguity (AI)} & \textbf{Risk} & \textbf{Risk (AI)} \\
\midrule
Preference Data   & 1.279  & 1.620   & 0.539* & 0.566  \\
                        & [0.733, 2.232] & [0.715, 3.670] & [0.308, 0.941] & [0.260, 1.232] \\
Sensitive Data     & 1.514  & 2.160  & 0.874   & 0.818  \\
                        & [0.863, 2.655] & [0.876, 5.326] & [0.502, 1.524] & [0.379, 1.767] \\
Constant (baseline odds)& 1.125  & 1.111   & 1.525* & 1.700 \\
                        & [0.763, 1.660] & [0.661, 1.869] & [1.024, 2.272] & [0.979, 2.953] \\
\midrule
Observations            & 302    & 133     & 308     & 160 \\
\bottomrule
\multicolumn{5}{l}{\footnotesize Odds ratios with 95\% CIs in brackets. Baseline framing: \neutral{}.}\\
\multicolumn{5}{l}{\footnotesize * $p<0.05$, ** $p<0.01$.}
\end{tabular}
\end{table}

\clearpage

\begin{figure*}[!htbp]
    \centering
    \includegraphics[width=0.7\linewidth]{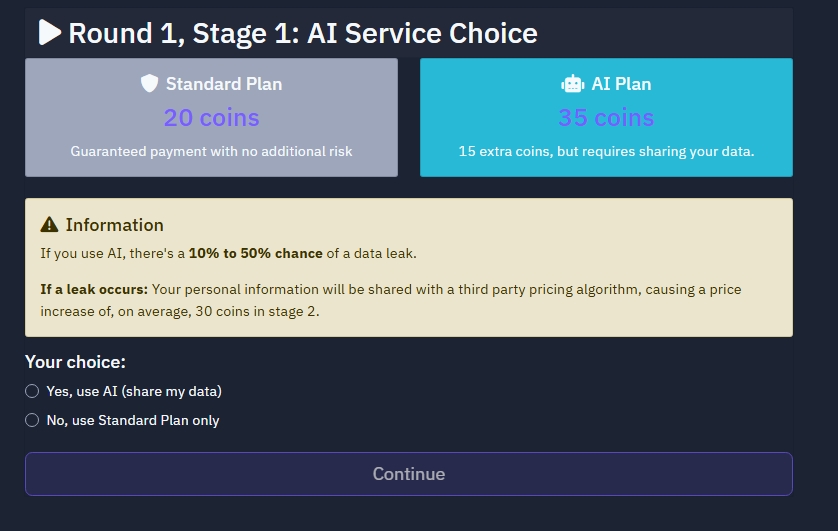}
    \caption{Main AI personalization decision screen before clicking on the personalized basket.}
    \label{fig:screen1}
\end{figure*}

\begin{figure*}[!htbp]
    \centering
    \includegraphics[width=0.7\linewidth]{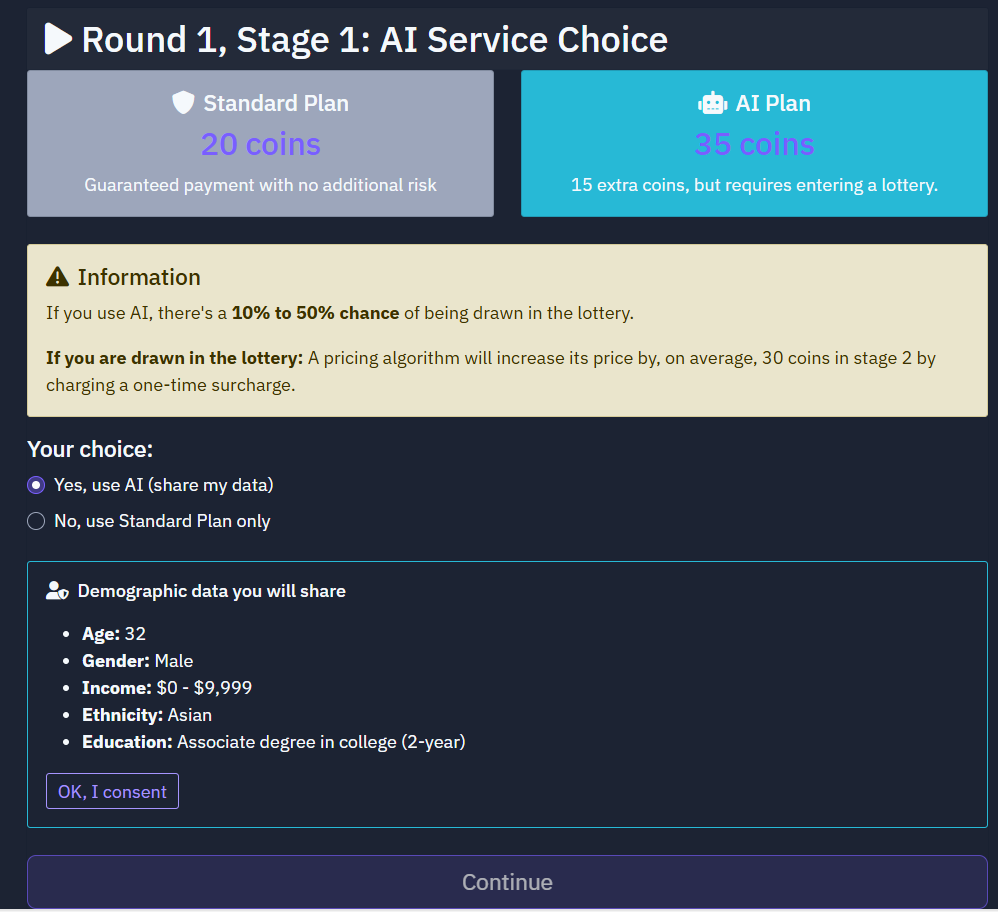}
    \caption{Main AI personalization decision screen after clicking on the personalized basket.}
    \label{fig:screen2}
\end{figure*}

\begin{figure*}[!htbp]
    \centering
    \includegraphics[width=0.7\linewidth]{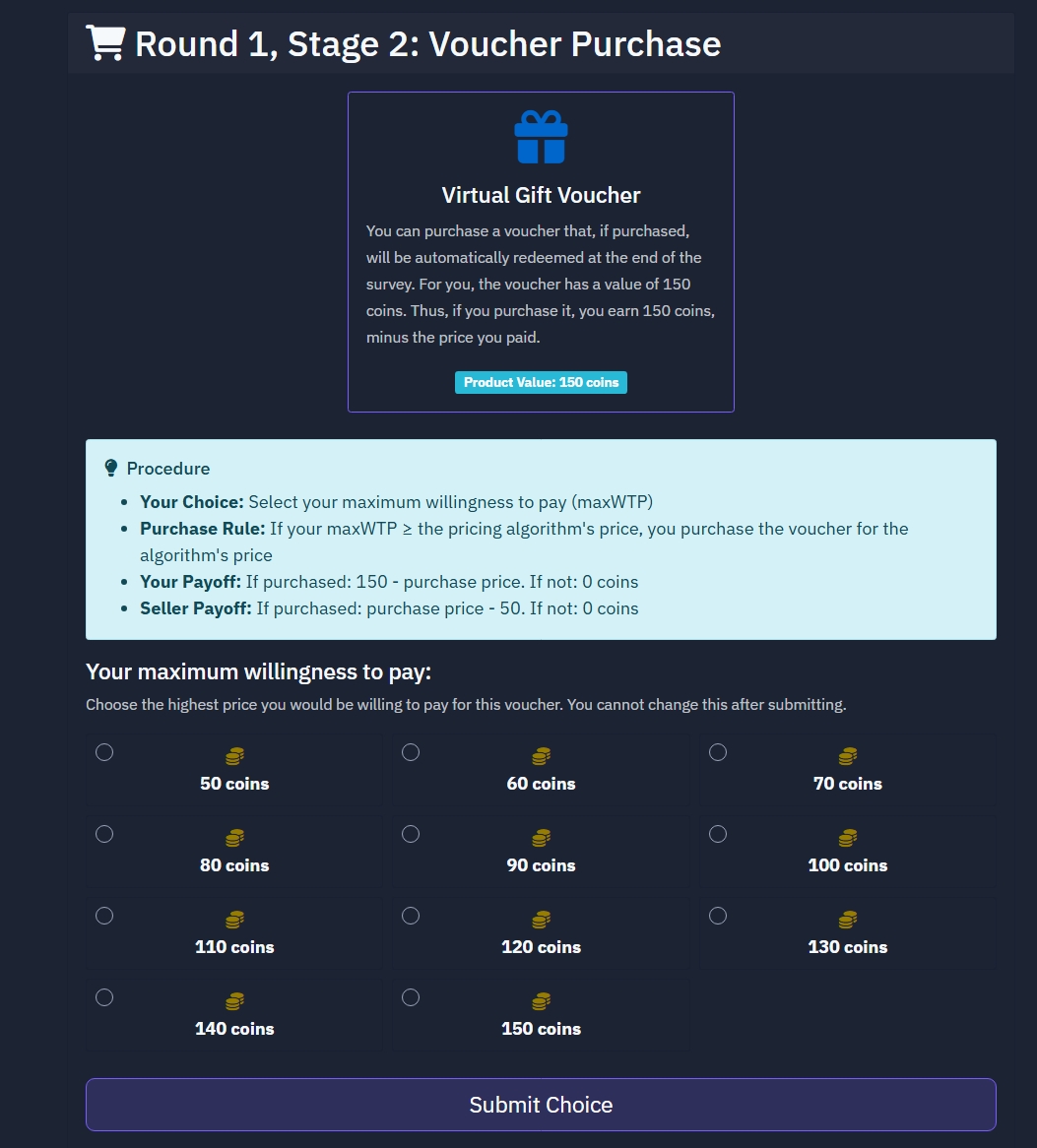}
    \caption{Ultimatum Bargaining Screen.}
    \label{fig:screen3}
\end{figure*}

\begin{figure*}[!htbp]
    \centering
    \includegraphics[width=0.7\linewidth]{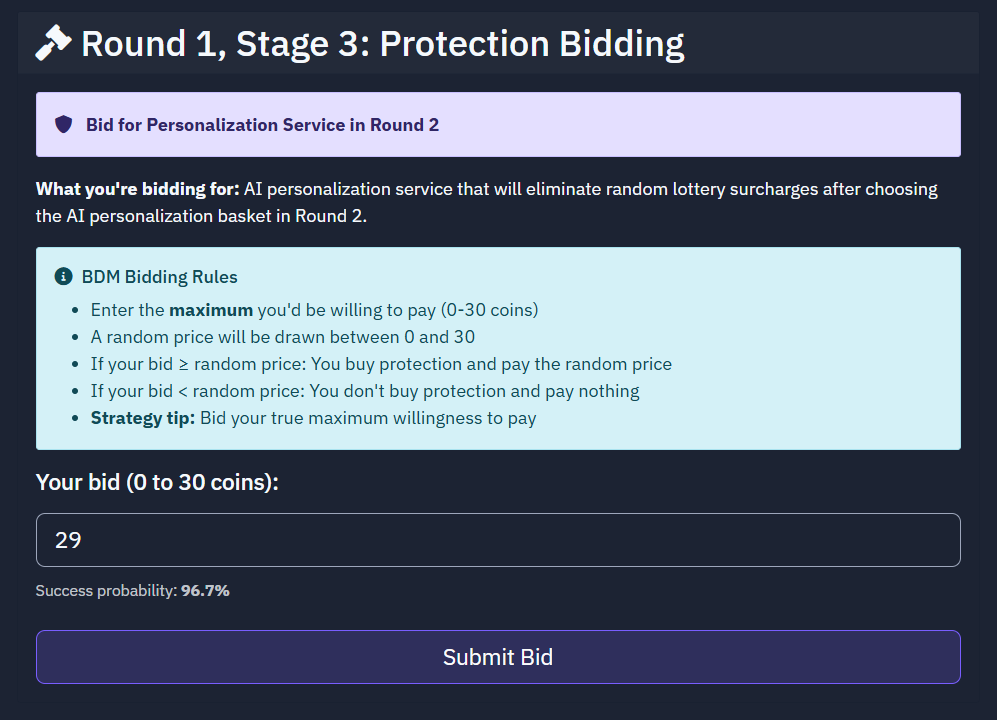}
    \caption{Privacy Label WTP Screen.}
    \label{fig:screen4}
\end{figure*}

\end{document}